\documentclass[fleqn,usenatbib]{mnras}

\usepackage{newtxtext,newtxmath}

\usepackage[T1]{fontenc}

\DeclareRobustCommand{\VAN}[3]{#2}
\let\VANthebibliography\thebibliography
\def\thebibliography{\DeclareRobustCommand{\VAN}[3]{##3}\VANthebibliography}

\usepackage{graphicx}	
\usepackage{amsmath}	

\usepackage{ulem}
\usepackage[dvipsnames]{color}

\def\be{\begin{equation}}
\def\ee{\end{equation}}
\def\ba{\begin{eqnarray}}
\def\ea{\end{eqnarray}}
\def\de{\partial}
\def\msun{M_\odot}
\def\div{\nabla\cdot}

\def\ltsima{$\; \buildrel < \over \sim \;$}
\def\simlt{\lower.5ex\hbox{\ltsima}}
\def\gtsima{$\; \buildrel > \over \sim \;$}
\def\simgt{\lower.5ex\hbox{\gtsima}}
\def\etal{{et al.\ }}

\definecolor{webgreen}{rgb}{0,.5,0}
\definecolor{webbrown}{rgb}{.6,0,0}
\definecolor{falured}{rgb}{0.5, 0.09, 0.09}

\title[Luminous disc-halo outflows]{Disc-halo gas outflows driven by stellar clusters as seen in multiwavelength tracers}

\author[E. O. Vasiliev \etal]{
Evgenii O. Vasiliev,$^{1}$\thanks{E-mail: eugstar@mail.ru}
Sergey A. Drozdov,$^{1}$\thanks{E-mail: sai.drozdov@gmail.com}
Biman B. Nath,$^{2}$\thanks{E-mail: biman@rri.res.in}
Ralf-J\"urgen Dettmar,$^{3}$\thanks{E-mail: dettmar@astro.rub.de}
Yuri A. Shchekinov,$^{1,2}$\thanks{E-mail: yus@asc.rssi.ru}
\\
$^{1}$Lebedev Physical Institute of Russian Academy of Sciences, 53 Leninskiy Ave., 119991, Moscow, Russia \\
$^{2}$Raman Research Institute, C. V. Raman ave., Sadashiva Nagar, Bangalore 560080, India \\
$^{3}$Ruhr University Bochum, Faculty of Physics and Astronomy, Astronomical Institute (AIRUB), Universit\"atsstrasse 150, 44780 Bochum, Germany
}

\date{Accepted XXX. Received YYY; in original form ZZZ}

\pubyear{2022}

\begin{document}
\label{firstpage}
\pagerange{\pageref{firstpage}--\pageref{lastpage}}
\maketitle

\begin{abstract}
We consider the dynamics of and emission from growing superbubbles in a stratified interstellar gaseous disc driven by energy release from supernovae explosions in stellar clusters with {masses $M_{cl}= 10^5-1.6\times 10^6~\msun$}. Supernovae are spread randomly within a sphere of $r_c=60$ pc, and inject energy episodically with a specific rate $1/130~\msun^{-1}$ proportional to the star formation rate (SFR) in the  cluster. Models are run for several values of SFR in the range $0.01$ to $0.1~\msun$ yr$^{-1}$, with the corresponding average surface energy input rate $\sim 0.04-0.4$ erg cm$^{-2}$ s$^{-1}$. We find that the discrete energy injection by isolated SNe are more efficient in blowing superbubbles: asymptotically they reach heights of up to 3 to 16 kpc for $M_{cl}=10^5-1.6\times 10^5~\msun$, correspondingly, and stay filled with a hot and dilute plasma for at least 30 Myr. During this time they emit X-ray, H$\alpha$ and dust infrared emission. X-ray liminosities $L_X\propto {\rm SFR}^{3/5}$ that we derive here are consistent with observations in star-forming galaxies. Even though dust particles of small sizes $a\leq 0.03~\mu$m are sputtered in the interior of bubbles, larger grains still contribute considerably ensuring the bubble luminosity $L_{\rm IR}/{\rm SFR}\sim 5\times 10^7  L_\odot M_\odot^{-1} ~{\rm yr}$. It is shown that the origin of the North Polar Spur in the Milky Way can be connected with activity of a cluster with the stellar mass of $\sim 10^5~\msun$ and the ${\rm SFR}\sim 0.1~\msun$ yr$^{-1}$ some 25--30 Myr ago. Extended luminous haloes observed in edge-on galaxies (NGC 891 as an example) can be maintained by disc spread stellar clusters of smaller masses $M_\ast\simlt 10^5~\msun$.  
\end{abstract}

\begin{keywords}
galaxies: halos -- starburst -- ISM: supernova remnants -- shock waves -- X-rays: galaxies -- infrared: galaxies
\end{keywords}



\section{Introduction} 

The discovery and intensive study of absorptions from heavy elements (such as CIV, SiIV, NV, OVI) in the circumgalactic medium (CGM) in quasar absorption spectra \citep[e. g.][]{proch06,simco06,pro17}, have posed stellar feedback as one of the most important physical factors that determines, along with gravity, the structure and evolution of galaxies. It {has} become clear that energy and mass exchange between interstellar discs and circumgalactic environment driven by energy release from massive stars and supernovae, is ubiquitous among galaxies. {Even dwarf galaxies at the low end of SFR reveal extended metal polluted haloes around them \citep{bur15,bord}. More recently, \citet{keeney} reported about presence of such haloes around galaxies without currently ongoing star formation.} This  suggests a very efficient mass exchange between galactic discs where heavy elements are produced and their distant neighbourhood in the halo. 

A common understanding is that the mass exchange between galaxies and their extragalactic environments is maintained by energy release from an enhanced SF that drives galactic winds.  Galactic wind, a large scale gas outflow from star formation in galactic discs, is thought to be driven by energy injection from young stars and supernovae {in starbursts events} with a {surface} SFR exceeding a certain critical value. The threshold  for galactic winds driven by central starbursts is estimated of $\dot\varepsilon\sim 10$ erg~cm$^{-2}$~s$^{-1}$ \citep{Lehnert1996a,heckman00}, and for disc-halo circulation in edge-on galaxies $\dot\varepsilon\sim 10^{-3}-10^{-4}$ erg~cm$^{-2}$~s$^{-1}$ \citep{dahlem,Rossa2000,Dahlem2006}. 
However, study of interrelations between the soft X-ray, UV, H$\alpha$, FIR and 1.4 GHz radio continuum emissions in a larger sample of 23 edge-on-galaxies led only to put a lower limit on the surface SN energy input rate $\dot\varepsilon\geq 10^{-3}$ erg~cm$^{-2}$~s$^{-1}$ \citep{tuel}. 

From a theoretical point of view, the evaluation of the energy threshold is a challenge. It is obvious that this threshold depends on many factors: ambient gas density, its vertical stratification -- characteristic scale heights and the circumgalactic floor density, concentration of energy sources both in space and time, dark matter distribution, energy injection regime and so forth  \citep{Rossa2000,Dahlem2001,Dahlem2006,Roy2013,nath13,vn15,Girichidis2016,vn17,Yadav2017,Fielding2018,vsn19}. Moreover, circulation of gas between different regions of a vertically stratified interstellar gas  requires different characteristic energy input rates. {From simulations} the required energy rate for driving low Mach number superbubbles ($M\simlt 3$) confined to the lower halo layer (within 2--3 scale heights) is $\dot\varepsilon\sim 10^{-4}$ erg~cm$^{-2}$~s$^{-1}$ \citep{vn17,Yadav2017,Fielding2018,Shchekinov2018,vsn19}, whereas a transition of gas circulation to the outer halo layers, being comparable in size to the galactic radial scale, seems to occur under shocks with higher Mach numbers ($M\simgt 10$) requiring an order of magnitude larger energy rate $\dot\varepsilon\simgt 10^{-3}$ erg~cm$^{-2}$~s$^{-1}$ \citep{Roy2013}. Early observations of our Galaxy towards the central region in 408 GHz have revealed loop-like structures -- Loop I amongst the most prominent \citep{Hanbury1960,berk,haslam}. More detailed analysis of the morphology of Loop I in 408 GHz and X-ray patterns from {\it ROSAT} led \citet{sofue} to conclude that Loop I is produced by a starburst in the Galactic center (radius of $\sim 200$ pc) with total energy of $E\sim 3\times 10^{56}$ erg within nearly 15 Myr (the equivalent SFR$\sim 2\msun$ yr$^{-1}$). The corresponding surface energy input rate lies in the range $\dot\varepsilon\sim 0.2$ erg~cm$^{-2}$~s$^{-1}$ derived in \citep{nath13}. A more recent numerical analysis has demonstrated that large scale structures similar to Loop I in our Galaxy might require even a higher energy input $\dot\varepsilon\sim 1$ erg~cm$^{-2}$~s$^{-1}$ \citep{sarkar15}. 

The discovery of the Fermi-Bubbles \citep{dob10,su10} has revealed a deficit of the SFR in the Galactic center (GC) as compared to the value needed for their maintenance. It was found that the GC shows  a modest level of star formation, which is  not sufficient for providing the energetics of Loop I \citep{sofue}. The estimates have been confirmed by more accurate evaluation from numerical simulations in \citet{sarkar15}. \citet{yusef} have inferred the history of SF rate in the GC (400 pc radius) from {\it Spitzer} and {\it Midcourse Space Experiment} and  concluded that the SF rate during the latest history -- the last 10 Myr, is only $\sim 0.04\hbox{--}0.08M_\odot$ yr$^{-1}$, while the SFR averaged over 10 Gyr is $\langle{\rm SFR}\rangle\simlt 0.14M_\odot$ yr$^{-1}$. The surface energy {input rate is thus $\dot\varepsilon\sim (2.5\hbox{--}5)\times 10^{-3}$ erg~cm$^{-2}$~s$^{-1}$ during the last 10 Myr, and $\langle\dot\varepsilon\rangle\sim 0.01$ erg~cm$^{-2}$~s$^{-1}$ over 10 Gyr.} Even though this number looks consistent to the limits determined in \citet{dahlem,tuel} for galactic discs, and the early estimate by \cite{heckman00} for galactic winds, {numerical simulations raise the lower limit of the required energy input by factor of at least 3--5 \citep{sofue,sarkar15}}. The inconsistency between the estimated SFR and the very existence of the Fermi-bubbles may be attributed to observational difficulties of inferring star formation rate on long time scales caused by a crowded environment in the Central molecular zone (CMZ), that causes uncertainties in counting stars of different age, measurements of infrared, bremsstrahlung emissions and other indicators of star formation. More recent discussions \citep[see, e.g.][]{barnes,feder,Krumholz2015,Krumholz2016,Kruijssen2017} {suggest that a short-term episodic regime of star formation with less pronounced observational manifestations can be possible.}

A similar phenomenon, the existence of extended haloes in galaxies with a relatively weak SF rate in the underlying discs, is observed in several edge-on galaxies in the local Universe. The galaxy NGC 891 represents a good example with the SFR$\approx 4\msun$ yr$^{-1}$ across the disc (optical radius $R_{25}\sim 20$ kpc) equivalent to $\dot\varepsilon\sim 10^{-4}$ erg~cm$^{-2}$~s$^{-1}$, and at the same time with a halo extending up to $\sim 2\hbox{--}5$ kpc in dust IR emission \citep[][see also a recently published catalog of dusty edge-on galaxies in \cite{Shinn2018}]{Howk1999,Alton2000,Rossa2004,Hughes2014,Seon2014,Bocchio2016,Yoon2020}, $\sim 5$ kpc in $2\hbox{--}5$ keV \citep{Hodges2018}, to $\sim 10$ kpc in soft X-ray ($0.4\hbox{--}1.4$ keV), and to $\sim 20$ kpc in HI \citep{Osterloo2007}. Extended circumgalactic gas traced by CIV, SiIV and OVI ions at projected radii of the order of 100 to 200 kpc are observed around galaxies often with a rather modest SFR$\sim 1\hbox{--}3\msun$ yr$^{-1}$ \citep[see discussion in][]{Tumlinson2011,bord}. Such extended haloes can either indicate that even a low SF rate under certain conditions is capable to drive circulation of gas within the inner and outer haloes as in the former case, or they are caused by powerful starbursts that have taken place in the past, e.g. 20--30 Myr ago, as in the latter.        

In this paper we focus on the ability of SF in galactic stellar clusters with a low to modest SFR ($0.02-0.1~\msun$ yr$^{-1}$), and energy injection concentrated in a relatively small volume, to drive outflows between the disc and the halo, and on their observational manifestations in X-ray, optical  and FIR tracers. In Sec. \ref{mod} we describe the model we use in simulations, Sec \ref{res} presents the results: i) evolution of the bubble under the action of a cluster depending on its SFR in Sec \ref{bubevol}, including also ii) destruction of dust particles in Sec \ref{dustd}, iii) the bubble emission characteristics -- X-ray, dust far-infrared, and H$\alpha$, and their possible interrelations in Sec \ref{emiss}, in Sec \ref{glxs} we consider possible implications of our results for the Milky Way and edge-on galaxies with a focus on NGC 891 among them, Sec \ref{summ} summarizes the results. 


\section{Model description} \label{mod}

\subsection{Equilibrium} 

We carry out 3-D hydrodynamic simulations (Cartesian geometry) of SN explosions inside a  
massive stellar cluster located  in the  galactic center. 
We study the dynamics of a bubble expanding preferentially perpendicular to the disc. We consider the bubble evolution during a period approximately twice as the lifetime of a least massive SN progenitor with $M\sim 8~\msun$, i.e. $\sim 35$~Myr. The gaseous disc is set up to be initially in hydrostatic equilibrium in the gravitational potential of the dark matter (DM) halo and the stellar disc \citep[see e.g., ][]{avillez00,hill12,walsch15,li17,vsn19}. The $z$-component of the gravitational acceleration due to the dark mater halo is calculated from a Navarro-Frenk-White profile with the virial radius of the halo equal to 200~kpc and concentration parameter $c=12$. 

\subsection{Stellar disc}\label{sdisk}

The stellar disc is assumed to be self-gravitating with an isothermal velocity dispersion. The acceleration perpendicular to the disc is $g_*(z) = 2\pi G \Sigma_* {\rm tanh} (z/z_*)$, where $\Sigma_*$ and $z_*$ are the stellar surface density and the scale height of the stellar disc. We adopt $\Sigma_* = 200~\msun$/pc$^2$ and  $z_* = 0.3$~kpc. 

\subsection{Gaseous disc} \label{Gd}

\begin{figure}
\includegraphics[width=8.5cm]{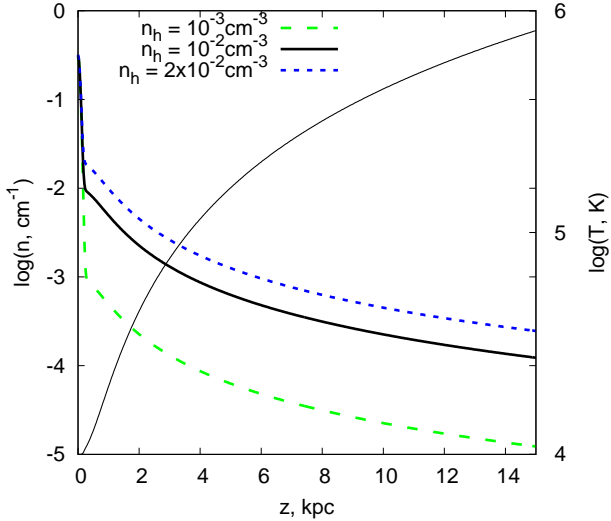}
\caption{
Gas density (left axis, thick lines) and temperature (right axis, thin line) profiles. The halo density is $n_h = 10^{-3}$ (dashed), $10^{-2}$ (solid) and $2\times 10^{-2}$~cm$^{-3}$ (dotted). The density at the disc midplane in all models is $n_d = 0.3$~cm$^{-3}$. The temperature profile is the same for all halo density values.
}
\label{fig-pro}
\end{figure}
 
The gas density in the midplane is assumed $n_0 = 0.3$~cm$^{-3}$ following \citep[][]{Kalberla2009}, which corresponds to gas surface density $\Sigma_{g} = 1~\msun$ pc$^{-2}$ with the scale height $z_g=0.2$ kpc at $R=1$ kpc. At larger heights above the midplane we assume a warm-hot gaseous halo with the profile $n(z) = n_h [1+(z/z_h)^2]^{-0.75}$, where $n_h = 0.01$~cm$^{-3}$, $z_h = 0.8$~kpc as follows from the MW halo distribution in \citet{miller15}. The temperature of the gaseous disc is set to $10^4$~K, the temperature of the halo increases with height to keep hydrostatic equilibrium. Figure~\ref{fig-pro} shows the initial gaseous density profiles in the vertical direction, with $n_h = 0.01$~cm$^{-3}$ as our  fiducial value. The initial metallicity in disc is assumed to be solar, in the vertical direction the metallicity decreases to its floor value [Z/H]$=-3$ at heights larger than 1~kpc. This allows us to better trace propagation of metallicity from exploded SNe into the halo. Initially the heating rate in the unperturbed disc and halo is set equal to the cooling rate according to the initial values of temperature and metallicity.
 
\subsection{Energy injection} \label{HN}

The dominant energy source feeding growing superbubbles is explosions of SNe, with a minor contribution from stellar wind as demonstrated recently by \citet{Franeck2022}. We inject the mass and energy by individual events with the energy corresponding to a joint energy of 30 standard SNe. Masses of standard SNe progenitors in a cluster are distributed randomly within 8--40~$\msun$ range according to the Salpeter initial mass function, the specific per mass SN rate is $\nu_m=1/130~\msun^{-1}$. Therefore, each joint injection carries the energy $3\times 10^{52}$~erg in thermal form and the mass load of 240-1200~$\msun$. This allows us to consider larger computational cells and assumes the injection radius corresponding to the Sedov-Taylor solution to be smaller than the cooling length. {The energy injection in form of thermal energy seems to be the most suitable in models with randomly spread in space and time point-like sources mimicing discrete SNe. As stressed above with the chosen cell size we avoid the overcooling problem discussed in \citet{Sharma2014}.} The interval between following energy injections corresponds in average to the lifetime of massive stars within 8--40~$\msun$ $t_{l} \sim M^{-1.57}$ \citep[see in][]{iben-book}. The energy and mass are injected randomly into a cell located within a spherical region of radius 60~pc centered at $(x,y,z)=(0,0,0)$. This procedure mimics random SNe explosions in a cluster. Overall, the energy injection rate in each of such clusters corresponds to explosions of standard SNe with the rate $\sim 10^{-4}-10^{-3}$ SN per year, or equivalently SFR$\sim 10^{-2}-0.1~\msun$ yr$^{-1}$, the corresponding stellar masses of clusters are in the high mass end of the clusters' mass function $M_{cl}\sim (2-16)\times 10^5~\msun$ \citep{cluster-rev}. Note that SFR$=1~\msun$ yr$^{-1}$ within the energy injection volume of radius $r_c=60$ pc corresponds to the surface SN energy injection rate $\epsilon\sim 2.5$ erg cm$^{-2}$ s$^{-1}$.

\subsection{The code and simulations} \label{tvd}
{
The code is based on the unsplit total variation diminishing (TVD) approach that provides high-resolution capturing of shocks and prevents unphysical oscillations. We have implemented the Monotonic Upstream-Centered Scheme for Conservation Laws (MUSCL)-Hancock scheme and the Haarten-Lax-van Leer-Contact (HLLC) method \citep[see e.g.][]{toro99} as an approximate Riemann solver. This code has successfully passed the whole set of tests proposed in \citet{klingenberg07}. 
}

Simulations are run with a radiative cooling described by a set of tabulated non-equilibrium cooling functions calculated by using the approach described in detail \citep{v13}. The functions are obtained for the gas cooling isochorically from $10^8$~K down to 10~K for metallicities within the range [Z/H]$=-4..1$. The non-equilibrium calculation \citep{v11,v13} includes kinetics of all ionization states of H, He, C, N, O, Ne, Mg, Si, Fe, as well as kinetics of molecular hydrogen at $T<10^4$~K. Fig.~\ref{fig-coolfun} in Appendix \ref{coolf} presents this set of cooling functions.

We apply a diffuse heating term representing the photoelectric heating of dust grains \citep{dust-heat}, which is thought to be the dominant heating mechanism in the interstellar medium. In our simulations, the heating rate is assumed to be time-independent and exponentially decreasing in the vertical direction with the scale height of the ISM disc. Such an assumption allows to stabilize radiative cooling of ambient gas with the temperature profile shown in Fig.~\ref{fig-pro}. Any deviation of the heating rate in the unperturbed gas violates the balance between cooling and heating and stimulates thermal instability, resulting in redistribution of gas mass in the interstellar disc \citep[see e.g. in][]{avillez00,hill12}. {The heating rate exponentially  decreasing upwards across the whole computational domain suppresses such contaminations \citep{li17}.}  

The simulations are performed with a physical cell size of 20~pc. The standard computational domain contains $768\times 384\times 768$ cells, that corresponds to $15.36\times 7.68\times 15.36$~kpc$^3$, but in several cases we extend the domain. We suppose a symmetry relative to the disc midplane and along the plane going through the cluster center perpendicular to the midplane, so we consider one fourth of the space. Usually we restrict our simulations within 35~Myr, in the end of this period the total vertical momentum of a bubble becomes negative. {We complete simulation earlier if the bubble reaches the borders of the computational domain, even though the total vertical momentum remains highly positive.} 


\section{Results}\label{res}

\subsection{Bubble evolution}\label{bubevol} 

\subsubsection{Gas}

\begin{figure*}
\center
\includegraphics[width=18.0cm]{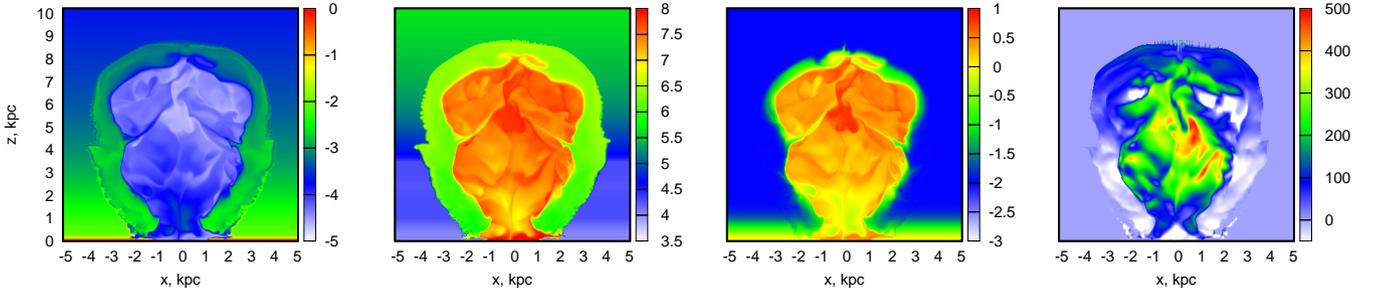}
\caption{
{
2D slices showing gas number density (log$[n$, cm$^{-3}]$, first panel), temperature (log$[T, K]$, second panel), metallicity ([Z/H], third panel), $z$-component of velocity ($v_z$, km s$^{-1}$, fourth panel) distributions in the outflow formed by SNe explosions in a cluster located at $(x,y,z)=(0,0,0)$ with SFR$=0.026~\msun$ yr$^{-1}$ ($\nu_{SN} \simeq 10^{-9}$~yr$^{-1}$~pc$^{-3}$). The bubble age is 25~Myr. The halo profile is fiducial with central density  $n_h = 10^{-2}$~cm$^{-3}$.
}
         }
\label{fig-2dslices}
\end{figure*}

In our model the cluster size is comparable to the disc scale height, therefore, the collective bubble formed by SNe with the rate $2\times 10^{-4}$ SN per year (SFR$=0.026~\msun$ yr$^{-1}$, $\nu_{SN} \simeq 10^{-9}$~yr$^{-1}$~pc$^{-3}$) considered here as an example, expands preferentially in the vertical direction, and during $\sim 1-2$~Myr reaches heights of $\sim 1-2$~kpc. When the shock wave approaches the interface between the disc and the halo at $z\sim 0.5$ kpc (seen in Fig.~\ref{fig-pro}) Rayleigh-Taylor instability breaks the shell and hot gas expands into halo with a larger scale height, resulting in a bottle-neck shaped bubble. At age $\sim 15$~Myr SNe explosions exhaust and the bubble continues expansion under momentum conservation. 

Figure~\ref{fig-2dslices} presents the hydrodynamical field in the outflow formed by {the cluster 
at the age 25~Myr -- left} to right: gas density, temperature, metallicity, $z$-component of velocity. The bubble reaches almost 9~kpc in height and about 8~kpc in its widest part. The bubble shell is still thick, around 1~kpc, and hot, $T\simgt 10^6$~K. Effects of radiative cooling are marginally observed in the lowermost region of the bubble edge in form of a small-scale density enhancement. {The bubble interior is hotter than its shell -- $T\simgt 10^7$~K, and will remain as hot further on {for a} characteristic cooling time $t_c\sim 30-100$~Myr. The bubble is still expanding with velocity $\simgt 100$~km s$^{-1}$, and with positive velocities dominating  almost in the entire bubble, except in the outer regions of the hot interior at heights $z\sim 6$ kpc and the compressed shell below $\sim 4$~kpc.} 

\subsubsection{Metals}
Metals ejected by SNe partly mix with the ambient hot bubble gas resulting in the metallicity spread from [Z/H]$=0$ to [Z/H]$=1$. It is worth to be pointed out that in our model mixing operates due to numerical diffusion $D_n\sim c_s\Delta x/3\sim 2\times 10^{26}$ cm$^{2}$ s$^{-1}$ in the shell with sound speed $c_s\sim 100$ km s$^{-1}$, and $\sim 10^{27}$ cm$^{2}$ s$^{-1}$ in the hot bubble with $c_s\sim 500$ km s$^{-1}$. This is consistent with the estimate of numerical kinematic viscosity $\langle\eta_n\rangle$ corresponding to the increase of specific per mass thermal energy $u$ as if it was grown due to viscousity at shock fronts $\langle\eta_n\rangle\sim \langle\dot u/(\div{\bf v})^2\rangle$. Within $\sim 25$ Myr numerical diffusion can mix metallicity over $\sqrt{\langle\Delta x^2\rangle}\sim 600$ pc which is half of the shell thickness. In the bubble interior mixing covers $\sim 3$ kpc in the same time range. In the thin external interface at the uppermost parts of the contact discontinuity it remains lower than [Z/H]$=-1$, indicating that mixing between the ejecta and ambient gas is slow. Inefficient mixing is one of the reasons that the shell still is hot, the cooling time in shell with [Z/H]$=-1$ is of the order 30--50 Myr. It is worth also noting that metals do not penetrate into the very external layer of the shell of $\sim 3$ kpc, being confined in a thinner layer $\sim 0.5-1$ kpc -- it can be seen when comparing the distributions of metals and temperature/density on Fig. \ref{fig-2dslices}. It is also seen that in the lower (conical) parts of the shell metals remain locked in even a thinner layer of $\sim 0.1-0.2$ kpc because of a lower temperature and higher density in this domain. At the asymptotic (``inertial'') stage with exhausted SNe explosions shown in Fig. \ref{fig-2dslices} metals are swept up from the superbubble centre, as can be observed in the central domain deficient in metals.

\subsubsection{Overall dynamics} 

\begin{figure}
\includegraphics[width=8.5cm]{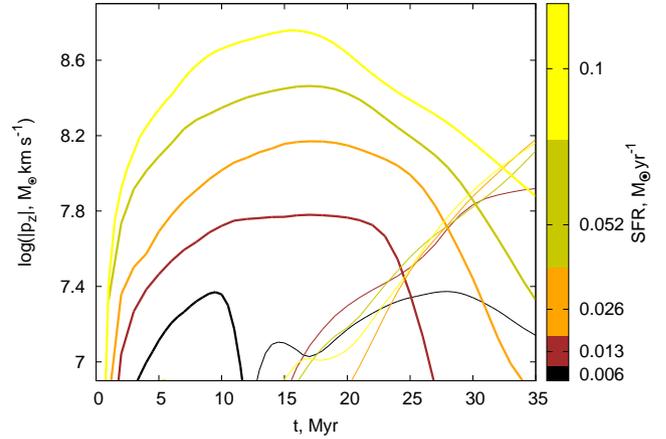}
\caption{
The $z$-component of momentum for outflowing ($p_z>0$, thick lines) and infalling ($p_z<0$, thin lines) gas flows for models with SFR=0.006, $0.013$, 0.026, 0.052 and 0.1~$\msun$ yr$^{-1}$.
}
\label{fig-evolp}
\end{figure}

\begin{figure}
\includegraphics[width=8.5cm]{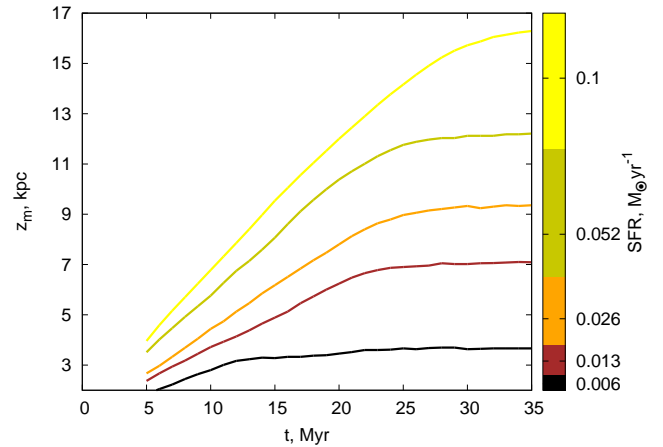}
\caption{
The height of a bubble versus the age for SFR=$0.006$, 0.013, 0.026, 0.052 and 0.1~$\msun$ yr$^{-1}$.        
}
\label{fig-evolh}
\end{figure}

Long-term evolution of superbubbles can be illustrated by time-behavior of their vertical momentum $p_z=\int\rho v_z dV$ shown in Fig.~\ref{fig-evolp} for several bubble models. Typically fast growth of $p_z(t)$ at very initial stages changes to a slow continuous gain on longer times $\sim 5-10$ Myr until reaching the maximum of $p_z$. For lower SFR, {except the lowest with the SFR$=0.006~\msun$ yr$^{-1}$}, the maximum stays on a plateau for long times (up to 15 Myr for SFR$=0.013~\msun$ yr$^{-1}$). On the contrary, superbubbles from clusters with higher SFR pass the maximum phase faster. The maximum and following drop of $p_z$ are connected mostly with a decrease of the shell mass because of a downward slipping of cold gas fragments along the shell. This can be the result of an enhanced pressure under more energetic shock waves from clusters with higher SFR, that stimulate faster gas cooling and the corresponding depletion of the shell mass. This process limits the expansion of shells. The superbubble from the minimum SFR$=0.006~\msun$ yr$^{-1}$ is an exception of this trend: the shell with such a low SFR expands on average into a denser environment than it gains for larger superbubbles, and its gas cools faster than gains energy from feeding explosions. The expansion phase ends by freezing out of the shells (Fig.S \ref{fig-evolh}), until they are disrupted by turbulence, by ongoing star formation nearby and/or galactic differential rotation. Depending on the SFR in the parent cluster it may take next tens of Myr as seen in Fig. \ref{fig-evolh}.  

The model with ${\rm SFR}=0.006~\msun$ yr$^{-1}$ seems to be close to the threshold SFR capable to drive the superbubles. Very rough estimates based on a comparison of the ram pressure from exploding SNe $L/4\pi z^2v_s$ on the shell of radius $R_s\sim z$, and the disc gravity $\rho_{sh}g_\ast z$, result in 
\be \label{lthre}
L\sim \pi\rho_{sh}g_\ast z^3v_s\sim 7.5\times 10^{41}nz_{3}^3v_{100} {\rm erg ~s^{-1}}
\ee 
for the disc parameters as in Sec. \ref{sdisk}. Here $L\sim \nu_{sn}E_{\rm B}$ is the mechanical luminosity of exploding SNe, $v_s$, the shell velocity $v_{100}v_s/(100~{\rm km~s^{-1}})$. The shell expansion velocity in the segments with predominantly vertical walls $v_{100}\simlt 0.3$, $z_{3}=z/3z_\ast$, and $g_\ast$ estimated at $z\simlt 3z_\ast$. With these assumptions Eq. (\ref{lthre}) is equivalent to the ${\rm SFR}\sim nz_{3}^3v_{100}~\msun$ yr$^{-1}$, or ${\rm SFR}\sim 0.001~\msun$ yr$^{-1}$ for $n\sim 10^{-3}$ cm$^{-3}$, $z=3z_\ast\sim 3$ kpc, and $v_s\sim 100$ km s$^{-1}$. It is important to stress though, that this conclusion is valid for star-formation localized in clusters with characteristic radius $R_c\sim 40$ pc, immersed into ambient gas with the parameters close to those defined in Sec. \ref{Gd}.  

\begin{figure*}
\center
\includegraphics[width=18.0cm]{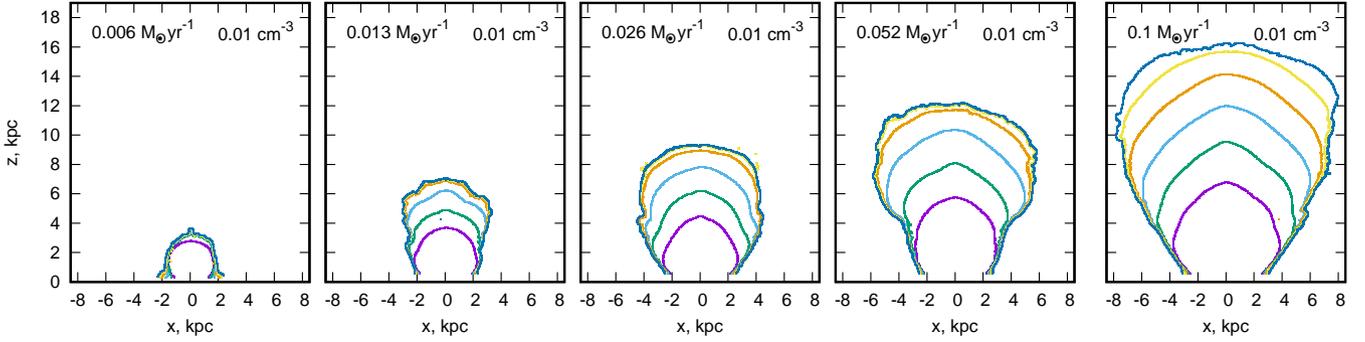}
\caption{
Left to right shown are contours of evolving bubbles with the SFR$=0.006, 0.013$, 0.026, 0.052 and 0,1~$\msun$ yr$^{-1}$. Contours depict the outer bubbles' edges at 10, 15, 20, 25, 30 and 35~Myr (from bottom to top). The halo profile is fiducial with $n_h = 10^{-2}$~cm$^{-3}$.
}
\label{fig-evolb}
\end{figure*}

Figure~\ref{fig-evolb} presents the outer borders (defined by the density jump) of the superbubbles evolving to this asymptotic: from left to right, we show the bubbles driven by clusters with SFR=0.006, 0.013, 0.026, 0.052 and 0.1$~\msun$ yr$^{-1}$ at several epochs. Nearly equidistant contours during the first 20~Myr show that the bubbles grow uniformly, i.e. with a constant velocity. After this period the bubbles driven by SFR$\sim 0.006-0.05~\msun$ yr$^{-1}$ start decelerating until their vertical expansion terminates by $t\sim 20-25$~Myr; the corresponding energy injection {rate $\dot\varepsilon\sim 0.0013-0.12$ erg~cm$^{-2}$~s$^{-1}$.} Bubbles of higher SF rates end their vertical expansion at later times. Radial expansion lasts several Myr shorter. The smallest superbubble in the first panel expands obviously slower as compared to those with a higher SFR. {Moreover, so far as in this case a considerable fraction of the superbubble expands into on average denser medium, the effects of radiative cooling in  the walls below $z\simlt z_h$, and the predominance of negative momentum from these regions, determine a fast decrease of the $z$-momentum component on Fig.~\ref{fig-evolp}.}  

\begin{figure}
\includegraphics[width=8.0cm]{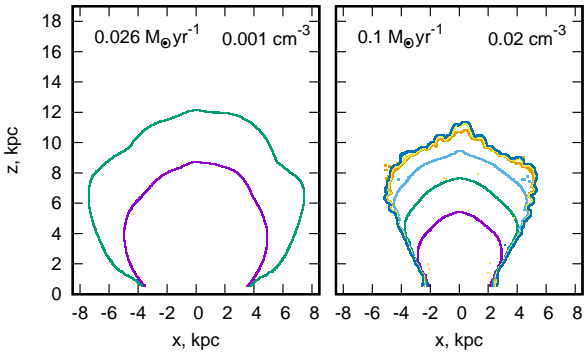}
\caption{
The two panels correspond to clusters with SFR$=0.026~\msun$ yr$^{-1}$ (left) and $0.1~\msun$ yr$^{-1}$ (right) immersed into a low ($n_h = 10^{-3}$~cm$^{-3}$) and a high halo densisty ($n_h = 2\times 10^{-2}$~cm$^{-3}$), respectively.
}
\label{fig-evolb2}
\end{figure}

The superbubble collimation increases with the halo gas density, as seen on two panels of Fig.~\ref{fig-evolb2}, because a higher density beyond the interface between the ISM disc and the halo suppresses the radial expansion.  

\subsubsection{Dust destruction}\label{dustd} 

\begin{figure}
\includegraphics[width=8.0cm]{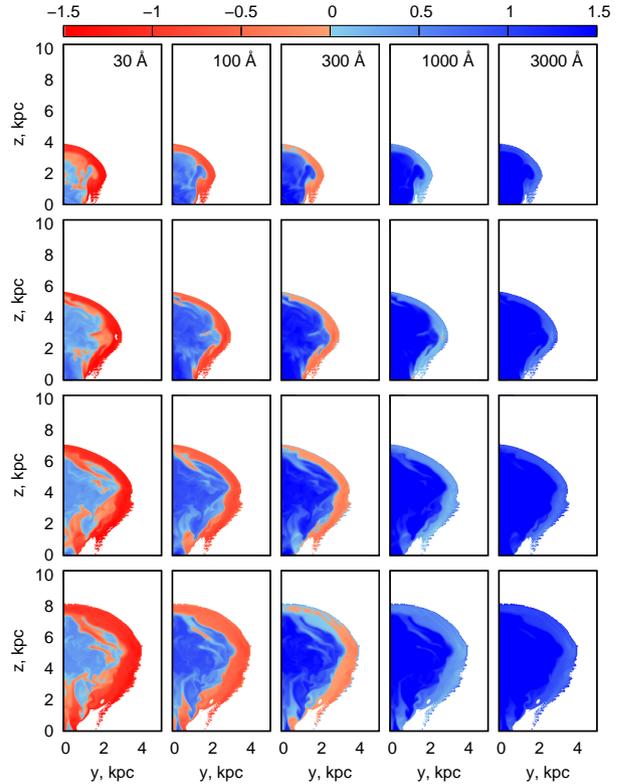}
\caption{
Distribution of the logarithm of the ratio $t_a(T)/ t_d$: left to right $a=30$,~100, 300, 1000 and 3000\AA. From top to boottom panels correspond to ages 10, 15, 20, 25 Myr.
         }
\label{fig-sput}
\end{figure}

\begin{figure*}
\center
\includegraphics[width=18cm]{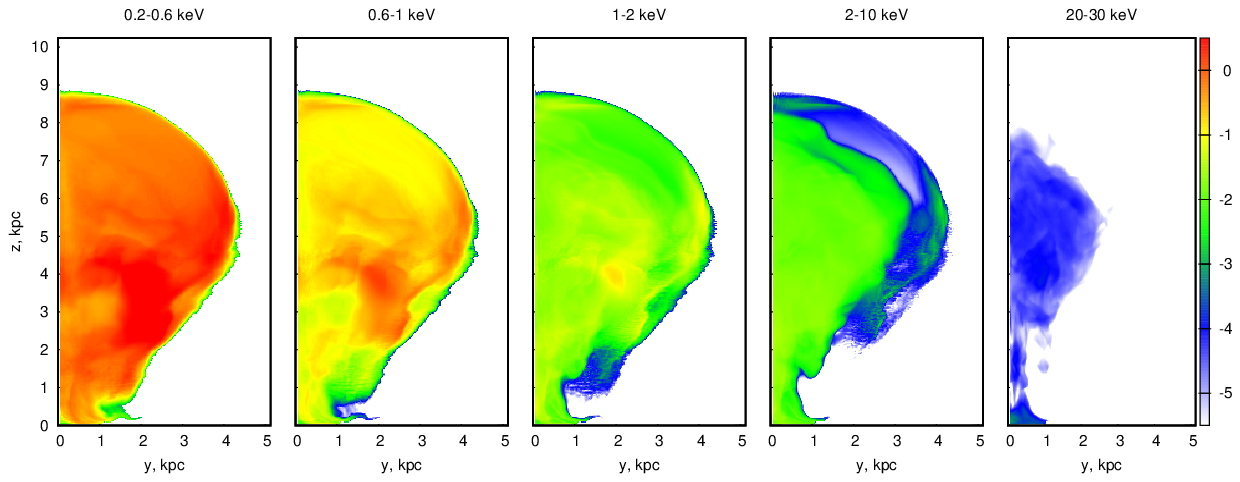}
\caption{
Edge-on ($yz$-plane) maps of the logarithm of X-ray surface intensity (keV~s$^{-1}$~cm$^{-2}$~sr$^{-1}$) in energy ranges 0.2-0.6, 0.6-1, 1-2, 2-10 and 20-30~keV (left to right) for the model with SFR=0.026~$\msun/$yr at the 25~Myr; the halo profile is fiducial: $n_h = 10^{-2}$~cm$^{-3}$. 
}
\label{fig-xraymaps}
\end{figure*}

Collisions of heavy ions of energies $E\simgt 30$ eV with dust particles result in their sputtering \citep{burke74,Draine1981,Dwek1981,Dwek1992}. The characteristic sputtering time in a given computational cell is approximated as $t_a(T)=a|da/dt|^{-1}\approx 10^5(1+T_6^{-3})a_{0.1}n^{-1}$ yr \citep[$T\simgt 10^6$ K, see Eq. 25.14 in ][]{draine-book}. In a magnetic free medium a nonthermal sputtering proceeds during the motion of a dust particle relative to plasma in a thin layer immediately behind the shock front $\Delta l\sim 10-20$ pc for the accepted parameters and thus can be neglected in our calculations. Thermal sputtering proceeds for the radiative cooling time $t_r\sim 10^5T_6n^{-1}$ yr unless local plasma is heated under the action of a shock from a distant SN. A rough estimate of time lags between two subsequent shocks impinging a given point in the bubble is $t_s\sim \tilde\nu_{sn}^{-1}$ varying from $2.5\times 10^2$ to $2\times 10^3$ yr depending on the SFR, here $\tilde\nu_{sn}=\int\nu_{SN}dV$ is the total SNR in the  cluster center. Such a high impinging rate prevents gas cooling and provides a continuous maintenance of high temperature. In this environment a dust grain loses around 90\% of its radius in $t=2t_a(T)$, and {hence in order to account dust destruction in a computational cell at a given stage the condition $t_a(T)\leq t_d$, with $t_d$ being the dynamical time, can be roughly accepted.} Therefore, as this condition is met, dust particles with radius $a$ are removed from a given computational cell. Figure \ref{fig-sput} presents the logarithm of the ratio $t_a(T)/t_d$ within the superbubble at different times for a set of particle radii.  

\begin{figure}
\includegraphics[width=8.5cm]{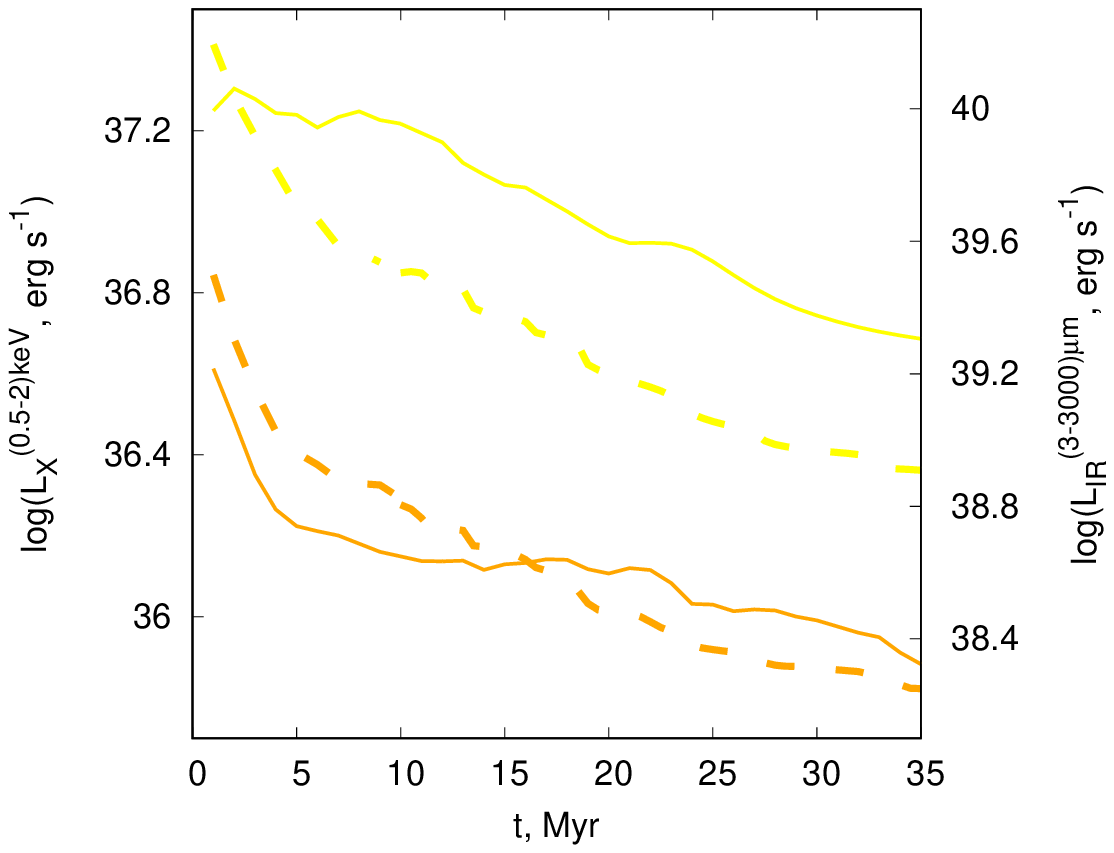}
\caption{
X-ray luminosity in energy range 0.5-2~keV (thin solid lines, left axis) and IR luminosity (thick dashed lines, right axis) from the whole bubble formed by  SFR=0.026~$\msun$ yr$^{-1}$ (two lower orange lines) and 0.1~$\msun$ yr$^{-1}$ (two upper yellow lines).
}
\label{fig-xraytot}
\end{figure}

As readily seen, smaller particles are sputtered first. However, the contribution of small grains to the thermal emission can remain considerable, particularly at higher frequencies because dust temperature $T_d$ is normally higher for smaller particles.  In order to estimate this effect we plot\footnote{We consider five bins of dust  with radius 30, 100, 300, 1000 and 3000\AA \ following to
the ``standard'' Mathis-Rumple-Nordsieck (MRN) distribution $n_a\propto a^{-3.5}$ \citep{Mathis1977}.} a distribution of the ratio $t_a(T)/t_d$ in the bubble over its evolution in Fig. \ref{fig-sput}. It is seen that during the whole evolution only small dust particles $a<300$\AA\, in the bubble shell are destroyed. In the low-density bubble interior the sputtering time is higher than $\sim 30\hbox{--}100~$Myr \citep{Dwek1992}, such that even small particles partly survive. Accounting for the fact that the shell sweeps up mostly the halo gas with a low dust mass fraction, the effects of dust destruction to the IR emission {is of minor importance}. 
 
\begin{figure}
\includegraphics[width=8.5cm]{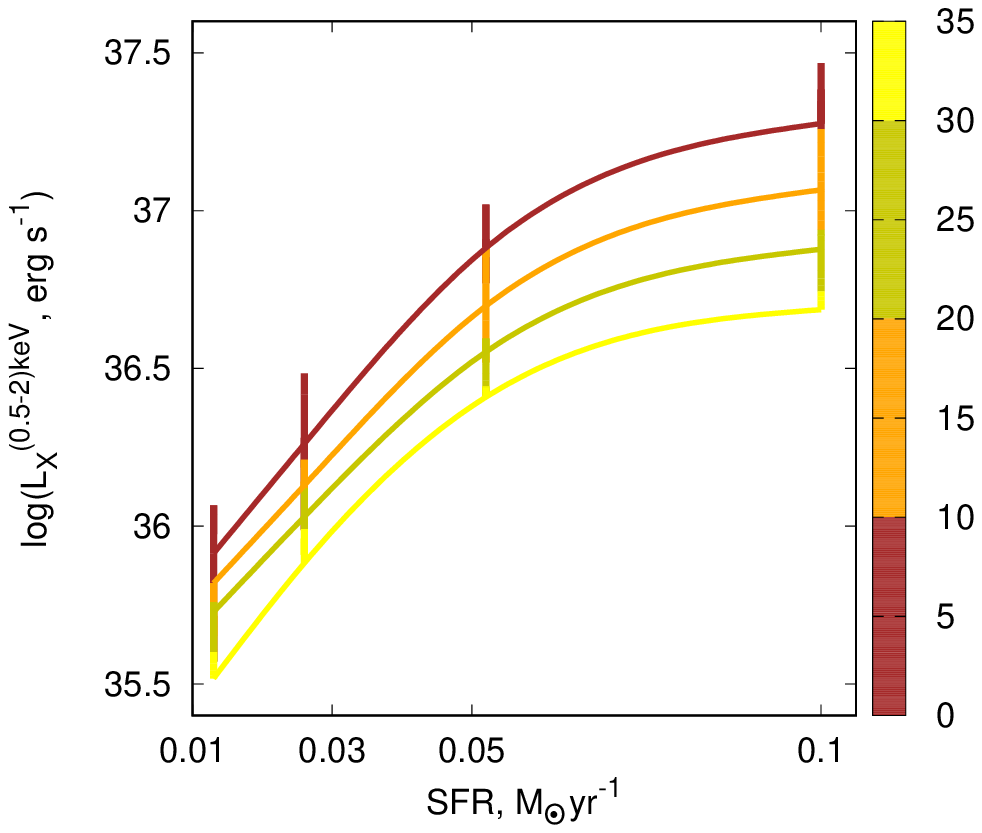}
\caption{
X-ray luminosity in energy range 0.5-2~keV from the whole bubble driven by SFR at 5, 15, 25 and 35~Myr (spline-smoothed lines from top to bottom). The vertical lines show evolution of $L_X$ for a given SFR value. The color bar encodes time in Myr.
}
\label{fig-xraysfr}
\end{figure}
 
Destruction of dust behind shock fronts is partly replenished by stellar activity via SNe factories  \citep{Todini2001,Bianchi2007,matsuura11,gall14,lau15}. We assume that besides the dust already present in the ISM gas, an additional dust mass is supplied into the bubble interior proportionally to the mass of ejected metals by SNe. {Collisional heating \citep{burke74,Draine1981,Dwek1981,Dwek1992} and destruction  \citep{Draine1979,Tielens1994,Jones1994} of the ejected dust is also implemented into our calculations.} 

\subsection{Bubble emission}\label{emiss}

\subsubsection{X-rays}\label{bubx}

\begin{figure}
\includegraphics[width=8.5cm]{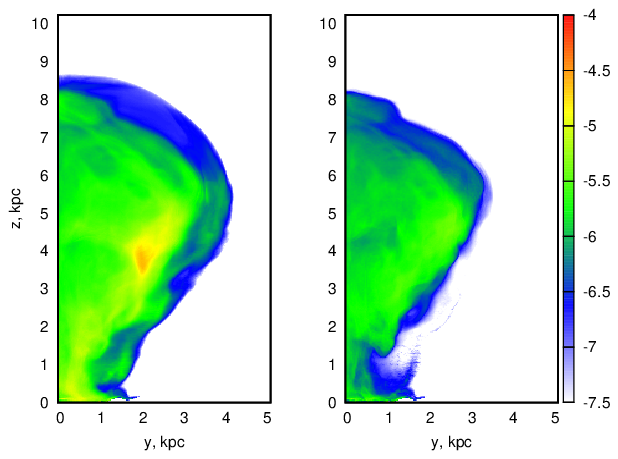}\\
\caption{The logarithm of the IR flux (in erg cm$^{-2}$ s$^{-1}$) from the dust immersed in the hot bubble: left panel -- ``normal'' dust size spectrum with $n_a\propto a^{-3.5}$ in the range $3\times 10^{-3}-0.3~\mu$m (dust is assumed to survive against sputtering), right panel shows the case for ``burnt'' dust predominantly with radii $0.03-0.3~\mu$m (presented by the sum of contributions from dust shown in the lower row in Fig. \ref{fig-sput}), SFR=0.026~$\msun$ yr$^{-1}$, the bubble age is 25~Myr.    
}
\label{fig-irmapup}
\end{figure}

Although radiative losses are too weak to affect the dynamics of the bubble during its expansion, the hot plasma in the interior emits sufficient energy in far-UV and X-ray ranges to manifest observationally. Figure~\ref{fig-xraymaps} presents the simulated X-ray emission (in log scale) of a bubble seen edge-on ($yz$-plane) at $t=25$~Myr, SFR=0.026~$\msun$ yr$^{-1}$. As seen, X-ray surface intensity of the bubble in the low-energy bands (0.2--0.6 and 0.6--1.0 keV) is considerable ($\sim 3$ keV s$^{-1}$ cm$^{-2}$ sr$^{-1}$) and comparable to that observed by eROSITA in the Galactic bubbles (see below in Sec. \ref{gbubbl}), and an order of magnitude lower than the intensity of the diffuse X-ray emission in the halo of NGC 891 galaxy \citep{ngc891xray}. It degrades slowly and stays approximately in the same range of intensity as shown in Fig.~\ref{fig-xraymaps} within next $\sim 30-40$ Myr, even after 25 Myr when SNe explosions exhaust. In general, the intensity is more or less uniform {throughout the bubble except several large-scale fluctuations in the highest energy bands close to the bubble border at heights $z\sim 2-8$~kpc.}

A slow cooling in the bubble interior keeps the gas hot and the total X-ray luminosity at a sufficiently high level. This is clearly seen in Fig.~\ref{fig-xraytot} where evolution of X-ray luminosity in the low energy band $0.5-2$~keV is shown. The luminosity in this band descreases gradually {roughly} as $L_X \sim t^{-1/4}$.

The total X-ray luminosity in the 0.5-2~keV band from the entire bubble increases for higher SFR approximately as $L_X\propto {\rm SFR}^{4/3}$ for ${\rm SFR}\simlt 0.06~\msun$ yr$^{-1}$, and flattens to $L_X\propto {\rm SFR}^{1/4}~\msun$ yr$^{-1}$ at higher SFR (Fig.~\ref{fig-xraysfr}). The flatness may be caused by the fact that for bubbles with ${\rm SFR}\simlt 0.06~\msun$ yr$^{-1}$ a higher fraction of the hot interior expands adiabatically. The `$L_X$-SFR' dependence in Fig. \ref{fig-xraysfr} is consistent with the integrated `$L_X$--SFR' relation ($E=0.3-2.0$ keV) for star-forming late-type edge-on galaxies \citep{tuel,Li2017}.

\subsubsection{IR dust emission} \label{IRdst}

\begin{figure}
\includegraphics[width=8.5cm]{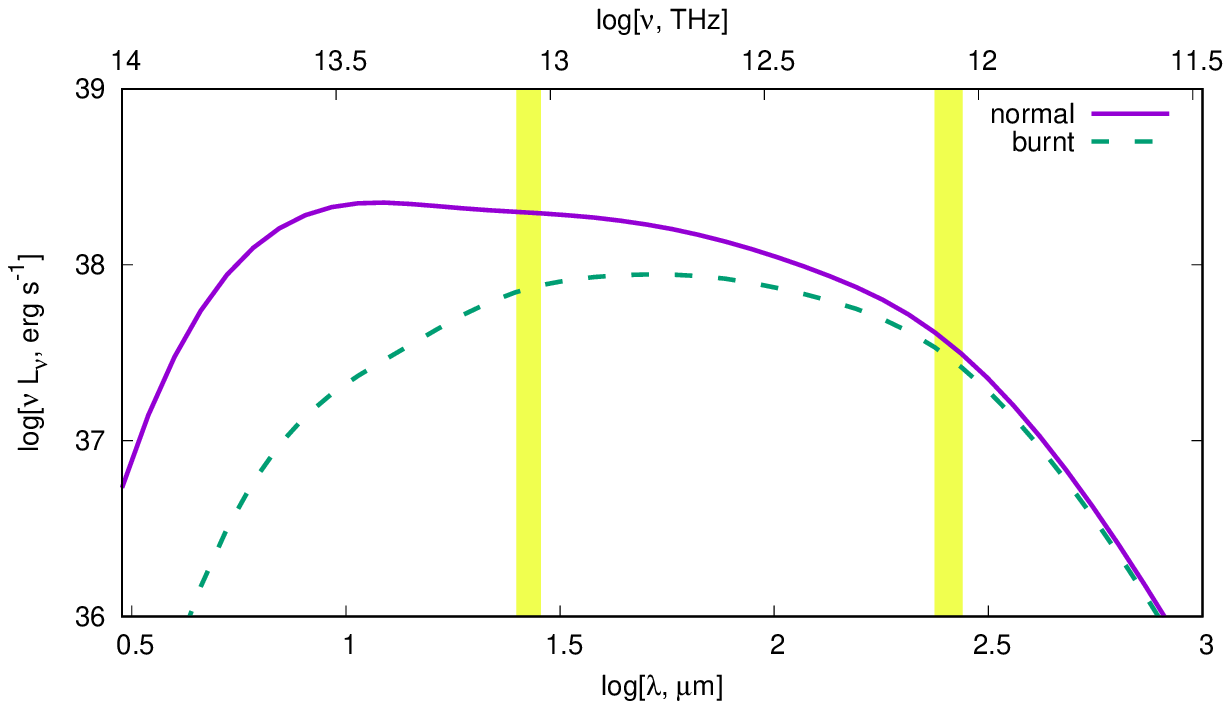}\\
\caption{
IR emission spectra of dust with ``normal'' and ``burnt'' dust composition: solid line represent the left panel of Fig. \ref{fig-irmapup} with ``normal'' dust, dashed line shows the spectrum of ``burnt'' dust (in the right panel of Fig. \ref{fig-irmapup}), the total (over the covered range) emission in the spectra relate as 0.35. The curves show the spectra {\it integrated} over all lines of sight in the entire bubble surface field, and thus represent the average spectrum.     
}
\label{fig-irspec}
\end{figure}

In order to evaluate the influence of dust destruction on their spectrum we calculate the dust emissivity with and without (depending on their lifetime) small dust grains -- {we refer the latter ``burnt'' dust}. The left panel in Fig.~\ref{fig-irmapup} presents the total IR flux (in log scale) produced by dust with sizes $a\sim 3\times 10^{-3}-0.3~\mu$m as if it was not destroyed in the entire bubble for the same model and age depicted in Figs.~\ref{fig-2dslices} and \ref{fig-xraymaps}. 
The flux of the dust emission in the right panel of Fig.~\ref{fig-irmapup} includes only grains with radii $a$ whose lifetime scale $t_a(T) \geq t_d$ as shown in lowest row Fig.~\ref{fig-sput}. In other words, in the bubble shell it mainly corresponds to the sum of contributions from dust particles $a=0.03~..~0.3~\mu$m as shown in the lowest row of Fig.~\ref{fig-sput}.
One can see that the smallest grains (30\AA) give a dim shell-like structure around a smooth bubble from larger grains at the level $\sim 3\times 10^{-6}~$erg~s$^{-1}$~cm$^{-2}$. Reduced FIR intensity in the shell on the right panel of Fig.~\ref{fig-irmapup} is obviously due to a deficient short wavelength emission from small-size particles. One can think that gas in the shells passed through hot and dense stages and destroyed small dust grains should manifest ``colder'' FIR spectra shifted towards longer wavelengths, in comparison with those  regions with ``normal'' dust size content, as seen on the right panel in Fig. \ref{fig-irmapup}. 
It is worth noting that even though sputtering of larger particles partly replenishes particles of smaller size, it is insufficient to compensate for their loss: first, because the sputtering time scales with radius $t_a\propto a$, and second, the number of larger particles is smaller, as for instance, in the case of $n_a\propto a^{-3.5}$ spectrum. Larger grains survive longer, as illustrated in Appendix \ref{sputter}.

\begin{figure}
\includegraphics[width=8.5cm]{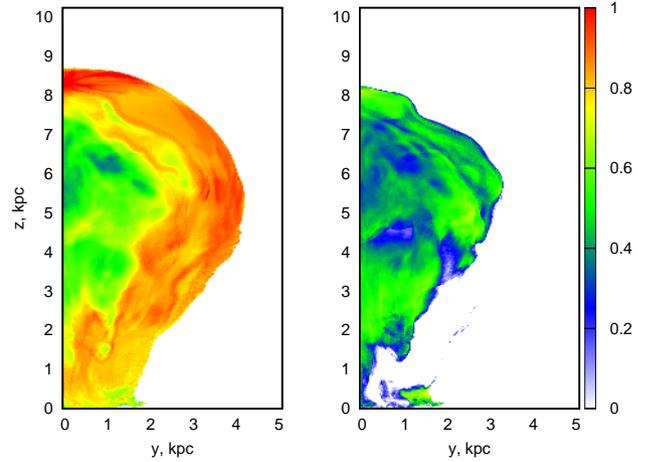}
\caption{The maps of distributions of the fluxes ratio (in log scale) of $F(25~\mu{\rm m})$ to $F_(250~\mu{\rm m})$ in the bands shadowed yellow in Fig. \ref{fig-irspec}. {\it Left}  and {\it right} panels reflect the ``normal'' and ``burnt'' dust as in Fig, \ref{fig-irmapup}.    
}
\label{fig-irmapcol}
\end{figure}

\begin{figure*}
\includegraphics[width=8cm]{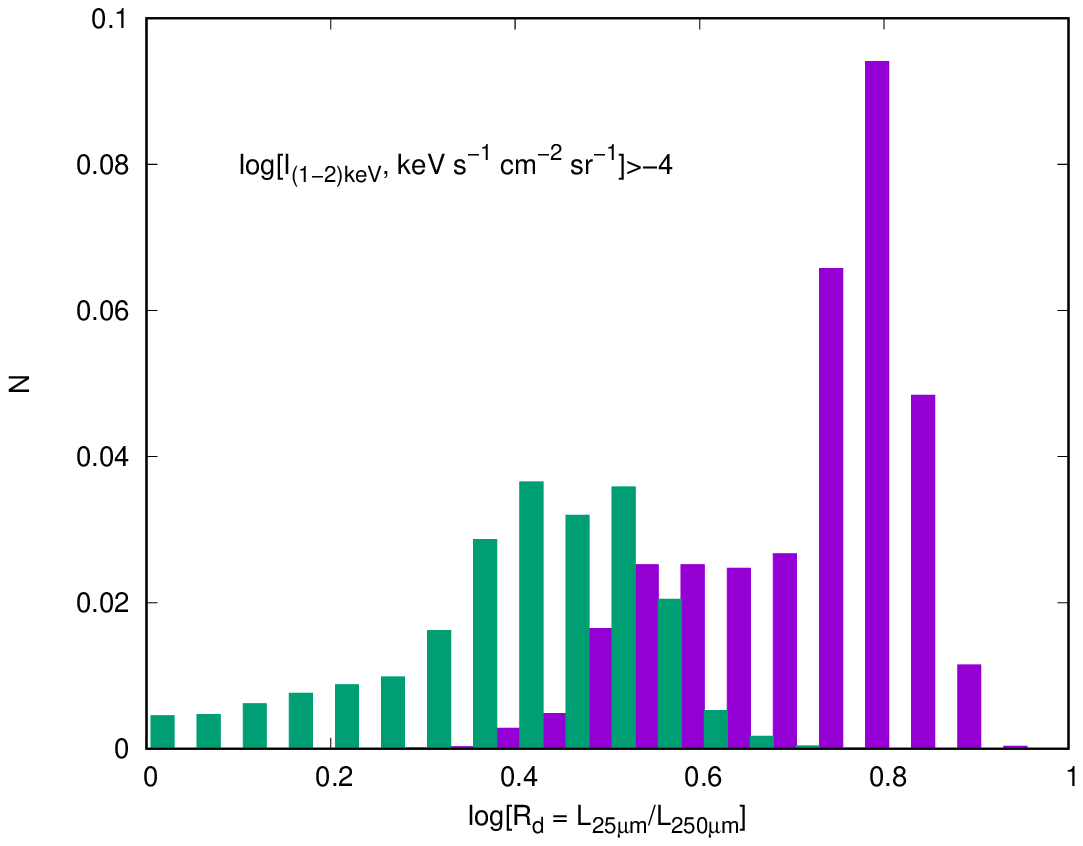}
\includegraphics[width=8cm]{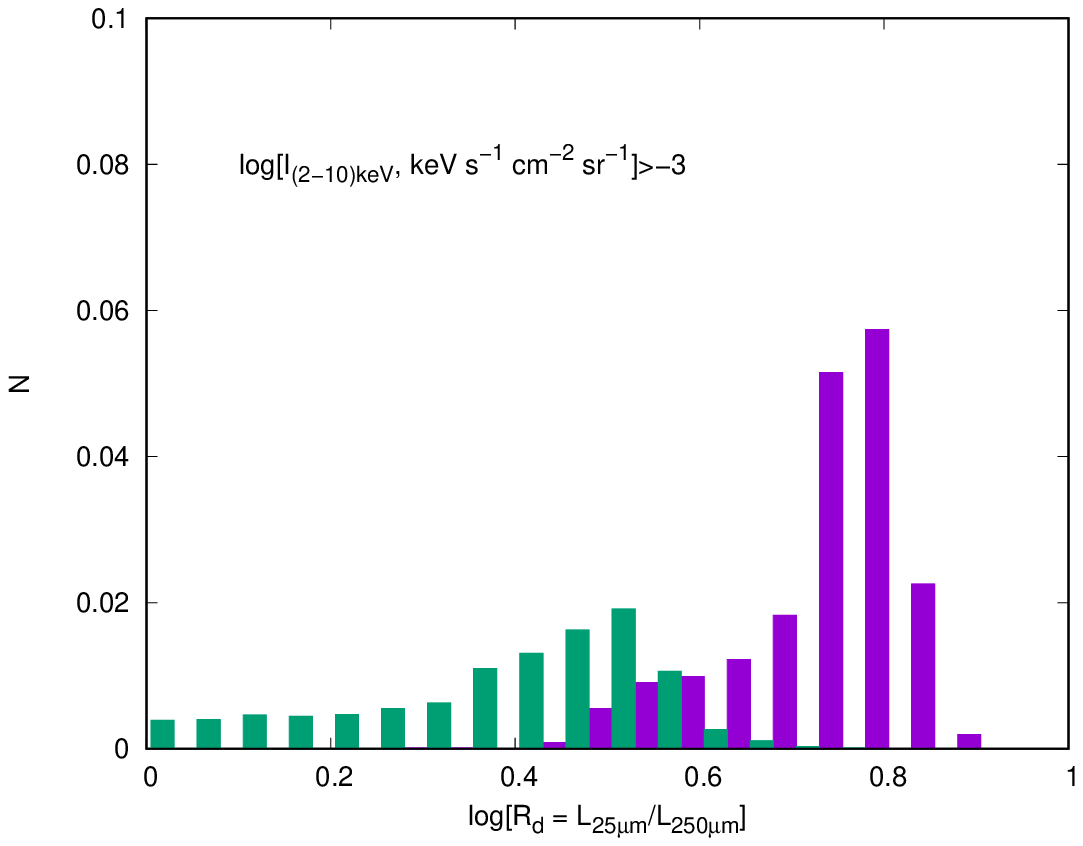}
\caption{{\it Left and right panels:} distributions of ``colors'' 25~$\mu$m/250~$\mu$m for a superbubble around the cluster with SFR$=0.026~\msun$ yr$^{-1}$ in regions radiating in the low $E=1-2$ keV and high $E=2-10$ keV energy bands. Magenta corresponds to a ``normal'', while green to a ``burnt'' dust. }    
\label{fig--ir-color}
\end{figure*}

As mentioned above, smaller particles have higher temperature and can contribute predominantly in emission spectra at higher frequencies. The latter is because that even in steady-state conditions dust temperature varies nearly as $T_d\propto a^{-1/(4+\beta)}$ ($\beta\simeq 2$ is the dust spectral index), whereas in hot dilute plasma stochastic heating from electrons makes small dust grains even ``hotter'' than at steady-state conditions \citep{Dwek1981,dwek86}: small dust particles experience higher magnitude of temperature fluctuations around a median value. At such circumstances, an ensemble of dust particles immersed in a hot environment show a thermal spectrum enhanced in the high frequency range. On the contrary, their deficit manifests a modified black-body spectrum with a monotonic 2nd derivative over frequency, as depicted by the dashed curve in Fig. \ref{fig-irspec}. {Observationally this difference can be obvisously recognized on ``color'' maps: the maps of the ratio $F_\lambda(25~\mu{\rm m})/F_\lambda(250~\mu{\rm m})$ -- the ratio of intensities at the wavelengths marked in Fig. \ref{fig-irspec} by yellow bands at which the contribution of small dust into the IR spectrum is clearly distinct. These maps for ``normal'' and ``burnt'' dust are shown on the left and right panels in Fig. \ref{fig-irmapcol}, corespondingly. The bubble with ``burnt'' dust is obviously seen to suffer of a deficit of emission at short wavelentgh as compared with the spectrum of an identical bubble with ``normal'' dust. The difference between the spectrum with a ``burnt'' and a ``normal'' dust can be also recognized in a correlation} between the dust FIR color, the ratio $F_\lambda(25~\mu{\rm m})/F_\lambda(250~\mu{\rm m})$ and the X-ray intensity. This correlation can be seen in the distribution function of the color over the superbubble edge-on projection as shown in Fig. \ref{fig--ir-color}: the ``burnt'' dust lacking of small size grains does not show the excess of emission at shorter wavelengths $\lambda<100~\mu$m, and thus has a lower ratio $F_\lambda(25~\mu{\rm m})/F_\lambda(250~\mu{\rm m})\simlt 0.5$, while the ``normal'' dust shows a higher ratio $>0.7$.  

\begin{figure*}
\includegraphics[width=18cm]{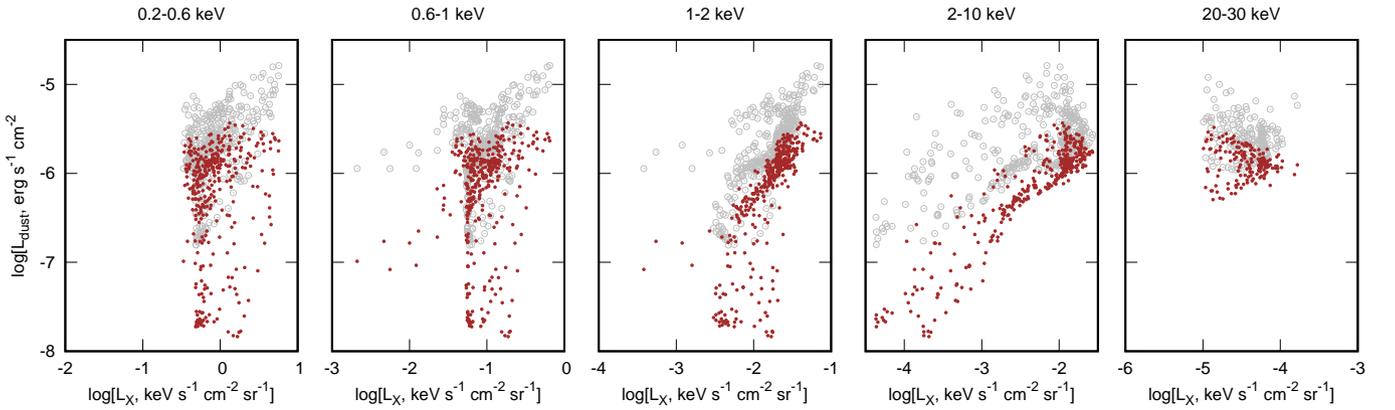}
\caption{
The interrelation between IR surface luminosity dust grains and X-ray intensity in energy  bands 0.2-0.6, 0.6-1, 1-2, 2-10 and 20-30~keV (left to right) for model with SFR=0.026~$\msun$ yr$^{-1}$. Large open (grey) circles present dust grains with ``normal'' dust-size composition $a=3\times 10^{-3}-0.3~\mu$m (presented on the left panel of Fig. \ref{fig-irmapup}), whereas small (dark red) dots correspond to the ``burnt'' particles with a narrower range of dust size-spectrum $a=10^{-2}-0.3~\mu$m (shown on the right panel of Fig. \ref{fig-irmapup}). As seen,  differences between the fluxes from ``normal'' and ``burnt'' dust is more that an order of magnitude in contrast to shown in Fig. \ref{fig-irspec}, as mentioned, this is because the spectra in Fig. \ref{fig-irspec} are averaged over the entire volume of the bubbles.  
}
\label{fig--ir-xray}
\end{figure*}

{Scatter plots in Fig.~\ref{fig--ir-xray} show the interrelation between the IR and X-ray surface intensities} in different energy ranges for a bubble fed by SFR=0.026~$\msun$ yr$^{-1}$. A remarkable difference between IR distributions of ``normal'' and ``burnt'' dust composition is that the former lies at high intensity end $\simgt 3\times 10^{-6}$ erg s$^{-1}$ cm$^{-2}$, whereas the latter tends to lower intensity due to a deficit of small grains (see discussion in Sec. \ref{IRdst}). The ``burnt'' dust deficient of smaller grains locates predominantly in the shell and in the edge layer between the shell and the hot bubble, i.e., it is connected spatially with denser and colder plasma. As a result, the low-brightness dust emission is present on panels with the low-energy X-ray bands as they occupy the entire field of the bubble including the dense shell and the edge between the shell and hot bubble. The energy band $E=2-10$ keV represents the bubble hot interior and its colder shell (Fig. \ref{fig-xraymaps}). This is reflected in the low-brightness tail both in X-ray and IR. The highest energy band {emission ($E=20-30$ keV) concentrates towards the bubble center and has the lowest brightness, The  dust in this region mostly avoids destruction from the hostile environment and is on average brighter.}

One can note similar behavior of the X-ray brightness in the $0.5-2$~keV band (third panel in Fig.~\ref{fig--ir-xray}) and the IR surface brightness (fifth panel). As mentioned early the total X-ray and dust IR luminosities also  demonstrate similarity in time dependence: they both decay as $L_{IR} \sim L_{X}(0.5-2{\rm keV}) \sim t^{-1/4}$ (Fig.~\ref{fig-xraytot}).

\subsubsection{Emissions vs height} \label{emissions}

\begin{figure*}
\center
\includegraphics[width=18.0cm]{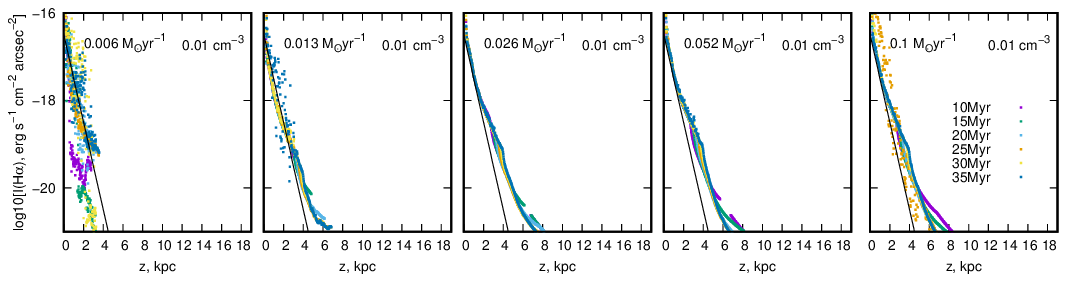}
\includegraphics[width=18.0cm]{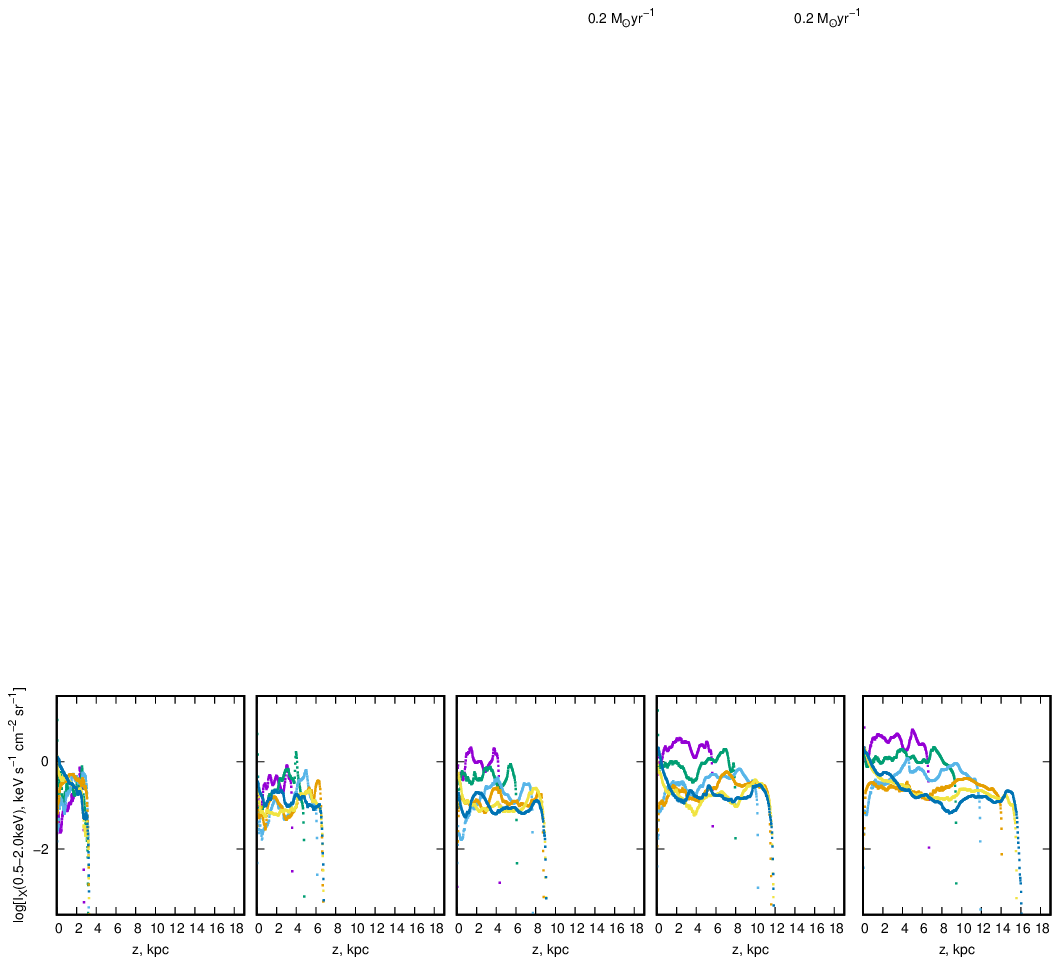}
\includegraphics[width=18.0cm]{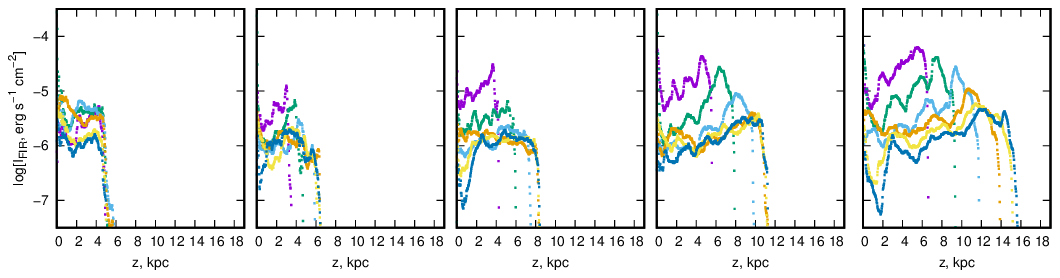} 
\caption{
{\it From the uppermost to the lowermost:} shown are the vertical profiles of H$\alpha$, X-ray (0.6-1~keV) intensities, and the IR flux in the range from $\lambda=3~\mu$m to $3=$ mm for different ages corresponding to the superbubbles in Fig. \ref{fig-evolb} (10, 15, 20, 25, 30 and 35~Myr, respectively, also shown in the upper right panel). The profiles are horizontally averaged at a given $z$ within $\Delta r=100$ pc centered at $r=0$. In panels of H$\alpha$ intensity the thin black line depicts an exponential decrease $I\sim {\rm exp}(-z)$. Colors code the superbubble age equivalently to those in Fig. \ref{fig-evolb}. 
}
\label{fig-emissions}
\end{figure*}

Emission characteristics of the superbubble interior depend on gas density and temperature, and therefore change in height and time. Thus they can characterize the bubble physical state and the age. Fig. \ref{fig-emissions} presents emission intensities along the vertical axis averaged over rings centered at $(x=0,y=0)$ for a given $z$-coordinate within radial direction $r = (x^2+y^2)^{1/2} = 100$~pc. These are: H$\alpha$ line, X-ray low-energy (0.6--1 keV), and $\lambda=3~\mu$m to $\lambda=3$ mm IR from dust, commonly used for diagnostics of gaseous haloes in edge-on galaxies.  

As can be seen, H$\alpha$ emission extends at a detectable limit $\simgt 10^{-19}$ erg s$^{-1}$ cm$^{-2}$ arcsec$^{-2}$ up to $z\sim 2$ kpc indicating that it comes from the lowest parts of the bubbles shells with a relatively high densities. X-ray low-energy band with small (within factor of 2) variations occupies the entire superbubble volume (as seen in Fig. \ref{fig-xraymaps}) and decreases in time. {IR dust emission is clearly seen to evolve similar to the low-energy X-ray emission. This reflects the fact that dust easily survives in the hot and X-ray bright bubbles' interior, as seen from comparison of Fig. \ref{fig-irmapup} with Fig. \ref{fig-xraymaps} (the first 3 panels). } 

{OVI ions are known to be very sensitive to temperature variations in a cooling plasma \citep{v11,v13}, and does not show a clear trend in the vertical distribution. However, it shows a rather intense 1036\AA\, line emission with $\simlt 10^{-18}$ erg cm$^{-1}$ s$^{-1}$ arcmin$^{-2}$ and can trace outflows of hot enriched gas into galactic haloes. }

\section{Implications: Milky Way and NGC 891}\label{glxs}

\subsection{The giant bubble in the Galaxy}\label{gbubbl}

{Estimates of SFR in the central Milky Way during recent 10 Myr vary around $\sim 0.04-0.08~\msun$~yr$^{-1}$ \citep{yusef,Immer2012,Koepferl2015,barnes} with the corresponding cluster mass of $M\sim 10^4-10^5\msun$, and as such can work as a source for keeping the central superbubble manifested in form of the North Polar Spur and Fermi-Bubbles.  Recently, eROSITA telescope installed onboard of the Spektr-RG mission has found an observational evidence of energy injection into the halo from past activity in the central part of the Milky Way \citep{erosita}. In the 0.6--1.0~keV band, the average observed X-ray intensity within the nothern and southern bubbles is around $\sim (2-4)\times 10^{-15}$~erg~s$^{-1}$~cm$^{-2}$~arcmin$^{-2}$. }

\begin{figure}
\includegraphics[width=8.5cm]{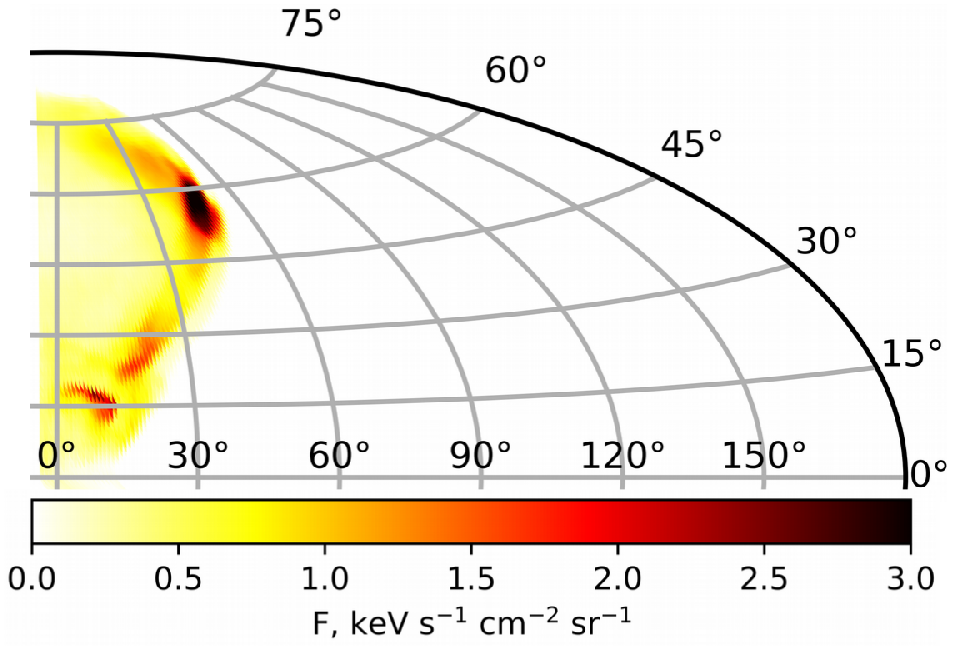}
\caption{
X-ray surface intensity (keV~s$^{-1}$~cm$^{-2}$~sr$^{-1}$) in the range 0.7-1.2~keV in the galactic coordinates (an observer is located at 8~kpc from the center of SNe cluster, which is in the galactic center) for the model with SFR=0.1~$\msun$ yr$^{-1}$. The age is 25~Myr. The halo profile is fiducial: $n_h = 10^{-2}$~cm$^{-3}$. 
}
\label{fig-molwp}
\end{figure}

In our simulations we consider the evolutionary and emitting properties of similar giant bubbles formed by SFR=$0.013-0.1~\msun$ yr$^{-1}$ (Fig.~\ref{fig-evolb}). The intensity obtained in our models is close to that  observed in the Galaxy. The bubble from a cluster with the SFR=$0.1~\msun$ yr$^{-1}$ reaches the height $\approx 14$~kpc at 25~Myr (forth panel on Fig.~\ref{fig-evolb}), that is close to the observed {height of the North Polar Spur}. For an observer at 8~kpc from the center the X-ray intensity (in 0.7-1.2~keV) of the bubble in galactic coordinates is shown in Fig.~\ref{fig-molwp}. The intensity in the bright shell is close to the values detected by eROSITA: $\sim (1-3)$~keV~s$^{-1}$~cm$^{-2}$~sr$^{-1}$. As can be seen from Fig. \ref{fig-evolh} at $t=25$ Myr the bubble continues to expand. In the next 10~Myr its flux decreases by a factor of $\sim 1.2-1.5$, and the shell rises upto latitudes of 80$^\circ$. The overall morphology is similar to the Milky Way bubble. Small-scale morphological differences are observed at lower latitudes, where details of inhomogeneous distribution of gas and star formation around the central galactic zone can introduce pecularities into dynamics. 

\citet{sarkar15} have described first a 2D dynamical model of a galactic scale outflow that {might have} resulted in formation of the Loop I structure and Fermi-bubbles inside. The overall dynamics is very similar to ours, with small differences in morphology, which are connected with the difference in energy injection regimes: spherically symmetric wind with a continuous rate in their model versus episodic explosions from individual SNe randomly spread in space and time within a central spherical stellar cluster of radius $r_c=60$ pc. The effective luminosities  though differ by factor of five: {star formation rates $0.5~\msun$ yr$^{-1}$ in \citet{sarkar15} model versus $0.1~\msun$ yr$^{-1}$ in our the most powerful model with the SFR close to the observed one} \citep{yusef,Immer2012,Koepferl2015,barnes}. The difference is also in total activity of the central source: \citet{sarkar15} considered the central source active for over 27 Myr, whereas in our case the star formation ceased at $t\sim 15$ Myr. The superbubble vertical scales are similar though: $\sim 15$ kpc at $t\sim 25-27$ Myr. The efficiencies of a continuous wind and sporadic individual SNe in driving large scale outflows are hard to compare because the models differ in their basic parameters: mid plane disc densities, halo densities and the corresponding scales. 

It is worth noting that the expected brightness of the superbubble in our model is comparable in order of magnitude to the one obtained by \citet{sarkar15}. For the leptonic mechanism of $\gamma$-emission connected with inverse scattering of CMB photons on cosmic ray electrons  with $E\sim 100$ TeV, one can estimated the surface brightness of the bubble at $\gamma$-ray energy $E_\gamma\sim 1$ GeV as $F_\gamma\sim 6\times 10^{-8}n_3T_7L_1$ GeV cm$^{-2}$ s$^{-1}$ sr$^{-1}$, with the gas density  $n_3=10^{-3}n$, and temperature $T_7=10^{-7}T$, and the line-of-sight thickness of the emitting hot gas $L_1=L/1~{\rm kpc}$, following the assumption of  \citet{sarkar15} that the cosmic ray electron energy density is proportional to the energy density of the hot gas.

\subsection{Edge-on galaxies} \label{fogs}

Our results can be applied also to edge-on galaxies where growing bubbles and superbubbles are seen projected along a sightline, and in some cases can be superimposed to form a smooth image along the galactic disc without being separated individually. 

The superbubble X-ray brightness in the low energy band $0.5-2$ keV band increases with the halo floor density $\rho_h$. This follows from the dependence of emissivity on density: the intensity is roughly $I_x\sim 2R_b\epsilon_X(\rho,T)$, where $\epsilon_X\propto\rho_b^2 T^{-1/2}e^{-1.16E_{\rm keV}/T_7}$ is the emissivity, $R_b\propto\rho_b^{-1/5}$ the bubble radius, $T\propto\rho_b^{-2/5}$. At later stages when the vertical size of the bubble is in excess of $\sim 2z_g$, it expands predominantly into the halo, and we assume that the mean gas density in the bubble is $\langle\rho_b\rangle\sim 10^{-2}\rho_h$ \citep{vn17}. The exponential factor in the emissivity $e^{-2.33/T_7}$ determines a slow decay of the brightness in time. 

Quite similarly H$\alpha$ emission increases with the halo density as the X-ray intensity does. It is connected with the fact that the recombination emissions come from the denser and colder shell with a dominant contribution from the regions belonging to the halo. 

For a fixed gas halo density profile the soft X-ray intensity of the bubbles shows {rather a weak sensitivity to SFR in the range $0.05-0.1~\msun$ yr$^{-1}$, as seen in Fig. \ref{fig-xraysfr} and on the second row in Fig. \ref{fig-emissions}:} from left to right the $z$-averaged intensity shows only time variations from $\sim 1-2$ at $t=10$ Myr to $\sim 0.1$ keV cm$^{-2}$ s$^{-1}$ sr$^{-1}$ at $t=35$ Myr. This is connected with a weak dependence of the superbubble radius on SFR nearly as $R_b\propto {\rm SFR}^{1/5}$. Another reason is that superbubbles with higher SFR expand progressively upwards after reaching a few $z_h$, and the gas density inside decreases faster than in those bubbles driven by lower SFR. However, when integrated over the superbubble the total X-ray luminosity relates to the SFR as $L_X\propto {\rm SFR}^\alpha$ with $\alpha\approx 0.25-1.3$ depending on the age (see, Sec. \ref{bubx}).   

Warm dust is also known to trace galactic outflows on scale heights from the inner and intermediate haloes \citep{Dahlem2001,tuel,McCormick2013} to larger heights reached by strong galactic winds. In the end of evolution the integral IR luminosity of a single bubble with a given SFR (in $\msun$ yr$^{-1}$) is $L_{\rm IR}/L_\odot\sim 2\times 10^6~{\rm SFR}$ (see Fig.~\ref{fig-xraytot}). This is strikingly lower (four order of magnitudes) than the $L_{FIR}$--${\rm SFR}$ relation for normal and starbursts galaxies \citep[see, for recent discussion][]{Kennicutt2012,Kennicutt2021}. It is important to emphasize {though} that this relation belongs solely to the superbubble interior extending to the outer $z>1$ kpc halo, and is explained by a low gas density in there. {On the contrary, FIR from starburst galaxies stems from the brightest regions of SF in the galaxy discs. } Estimates of the extraplanar ($z\simgt 1$ kpc) FIR emission in edge-on galaxies give similar numbers. For instance, the sum of the FIR fluxes from extraplanar regions around the X-shaped structure in the galaxy NGC 3079 with its SFR$\approx 2.6~\msun$yr$^{-1}$ \citep[][]{Veilleux2021}, results in $L_{FIR}\sim 2 \times 10^7~{\rm SFR}$ in solar units. A similar value $L_{FIR}\sim 4\times 10^7~{\rm SFR}$ can be inferred from the data for the halo of the galaxy NGC 891 presented by \citet[][see their Table 3]{Yoon2020}. 
 
\subsubsection{NGC 891} \label{ngc891}

The galaxy NGC 891 is similar to the Milky Way \citep[][]{Bok1981,Kruit1984,Sofue1987,Bottema1991,Dettmar1992,dahlem}. Its IR, H$\alpha$, CO, [CII]~158$\mu$m emissions and the synchrotron radio-halo and other galaxy-scale structures are fed by SFR$\sim 5$ to $\sim 10~\msun$ yr$^{-1}$ \citep[as inferred from][]{dahlem}. Recently \citet{Yoon2020} described a  FIR emission from a dusty halo extending upto 8 kpc above the disc. It may indicate that even though averaged over the disc surface SFR is only $\sim 0.03~\msun$ yr$^{-1}$ kpc$^{-2}$, {under certain conditions it can launch a large scale elevation of the interstellar gas.} From the point of view of our simulations, such conditions can be fulfilled when the injection of energy into the ISM is provided by compact stellar clusters: about a hundred OB-associations of $\sim 0.04~\msun$ yr$^{-1}$ with the total rate $\sim 4~\msun$ yr$^{-1}$ spread randomly over the disc can maintain such a halo. Indeed, as seen from Fig. \ref{fig-evolb}, a cluster with SFR$=0.04~\msun$ yr$^{-1}$ in 35 Myr reaches $z\sim 10$ kpc and $r\sim 4$ kpc, covering in the disc $\sim 50$ kpc$^2$. As seen in Fig. \ref{fig-evolh} the lifetime of such superbubbles can be longer than 35 Myr, resulting in a high covering fraction of the disc area by superbubbles. The total X-ray luminosity of such superbubbles can be of $10^{36}-10^{37}$ erg s$^{-1}$ each as can be judged from Fig. \ref{fig-xraytot}, being in agreement with observational estimates \citep{Bregman1994,Wang1995}. 

\citet{Hodges2018} report a disc-wide X-ray emission $E=0.4-1.4$ keV with the intensity $\sim 10^{-15}$ erg~cm$^{-2}$~s$^{-1}$~arcmin$^{-2}$ extending up to 5 kpc in the vertical direction, and up to 10 kpc in the central part; higher energy photons $E=2-5$ keV extend over smaller distances. All our models except the one with SFR$=0.026~\msun$ yr$^{-1}$ and $n_h=0.001$ cm$^{-3}$ (5th panel in Fig. \ref{fig-evolb}) predict similar intensities within the heights they occupy. In particular, the model superbubble from a cluster witn SFR$=0.05~\msun$ yr$^{-1}$ extends its soft X-ray up to $z\sim 10$~kpc (see third panel in second row in Fig.~\ref{fig-emissions}). The latter may indicate on the presence of an active stellar cluster that might has initiated an outflow 25--30 Myr ago in the center of NGC 891.

Intensities in H$\alpha$ predicted in all our models (except the one with SFR$=0.026~\msun$ yr$^{-1}$ and $n_h=0.001$ cm$^{-3}$ on 5th panel in Fig. \ref{fig-evolb}) within $|z|\sim 2-3$ kpc are consistent with those observed in edge-on galaxies in the sample reported by \citet{Miller2003} including NGC 891.

The OVI 1036\AA\, line emission with $\simlt 10^{-18}$ erg cm$^{-1}$ s$^{-1}$ arcmin$^{-2}$ at heights up to $z=1.5$ kpc is also observed \citep{Chung2021}. Absorptions in low-ionized ions typical for warm ionized gas (CII, MgII, SiII, SiIII and others) has been observed by \citet{Qu2019}.


\section{Conclusions} \label{summ}

Our results are summarized as follows:  

\begin{enumerate}
\item Stellar clusters {with SFR$=0.006-0.1~\msun$ yr$^{-1}$ and surface energy injection rate of $\dot\varepsilon\sim 0.01-0.15$~erg~cm$^{-2}$~s$^{-1}$ produce} superbubbles extending from 3 to 16 kpc. The superbubble traced in the Milky Way by the North Polar Spur can be launched and maintained by a cluster with a relatively low ${\rm SFR}\sim 0.1~\msun$ yr$^{-1}$ close to the observed one. {The value of the ${\rm SFR}\sim 0.006~\msun$ yr$^{-1}$ and the corresponding energy input rate $\dot\varepsilon\sim 0.01$ erg cm$^{-2}$ s$^{-1}$ is close to the threshold in ambient gas with parameters {typical for} the Milky Way central region. }

\item Such superbubbles can stay filled with a hot ($T\simgt 10^6-10^7$~K) low-density ($n \simlt 10^{-3}$~cm$^{-3}$) {and enriched (${\rm [Z/H]}\simgt 0$) gas} for a long time of the order of $t\simgt 30$ Myr, and be sufficiently {bright in the soft X-ray emission in energy range $E=0.5-2.0$ keV with $L_X\sim (0.1-1)\times 10^{36}~{\rm SFR}^{4/3}$ erg s$^{-1}$ while ${\rm SFR}\simlt 0.06~\msun$ yr$^{-1}$, and  $L_X\sim (2-6)\times 10^{37}~{\rm SFR}^{1/4}$ erg s$^{-1}$ at higher SFR. }

\item A fraction of dust can survive in the bubble interior on this time scale, and can be seen in infrared emission. However, a deficit of dust particles of small sizes $a\leq 100$ \AA\, results in a considerable, factor of ten, depression of the integrated dust IR emission, even though the dust mass remains practically unchanged. The deficit of small size particles is also revealed in the ratio $F_\lambda(25~\mu{\rm m})/F_\lambda(250~\mu{\rm m})$. The ``normal'' dust shows $\log[F_{25}/F_{250}]\approx 0.8$, whereas the ``burnt'' one has $\log[F_{25}/F_{250}]\approx 0.5$.
 

\item Stellar clusters with SFR in this range spread through over galactic discs can maintain gaseous haloes that radiates in X-ray, H$\alpha$, {and IR dust continuum with the intensities similar} to those observed in edge-on galaxies (such as NGC 891). The luminosity of individual supperbubbles is connected with the underlying SFR $L_{\rm IR}/L_\odot\sim 2\times 10^6~{\rm SFR}$, roughly consistent with the halo IR-emission observed in a few edge-on galaxies. 
\end{enumerate}
\par


\section*{Acknowledgements}

We thank the referee, R. W{\"u}nsch, for his friendly criticism and valuable comments. The work by SD and YS is done under partial support from the pro ``New Scientific Groups LPI'' 41-2020. The numerical simulations have been performed in the High Performance Cluster at the Raman Research Institute, Bangalore.

\section*{Data Availability}

The data underlying this article are available in the article.


\appendix

\section{Cooling function} \label{coolf}
\begin{figure}
\includegraphics[width=8.0cm]{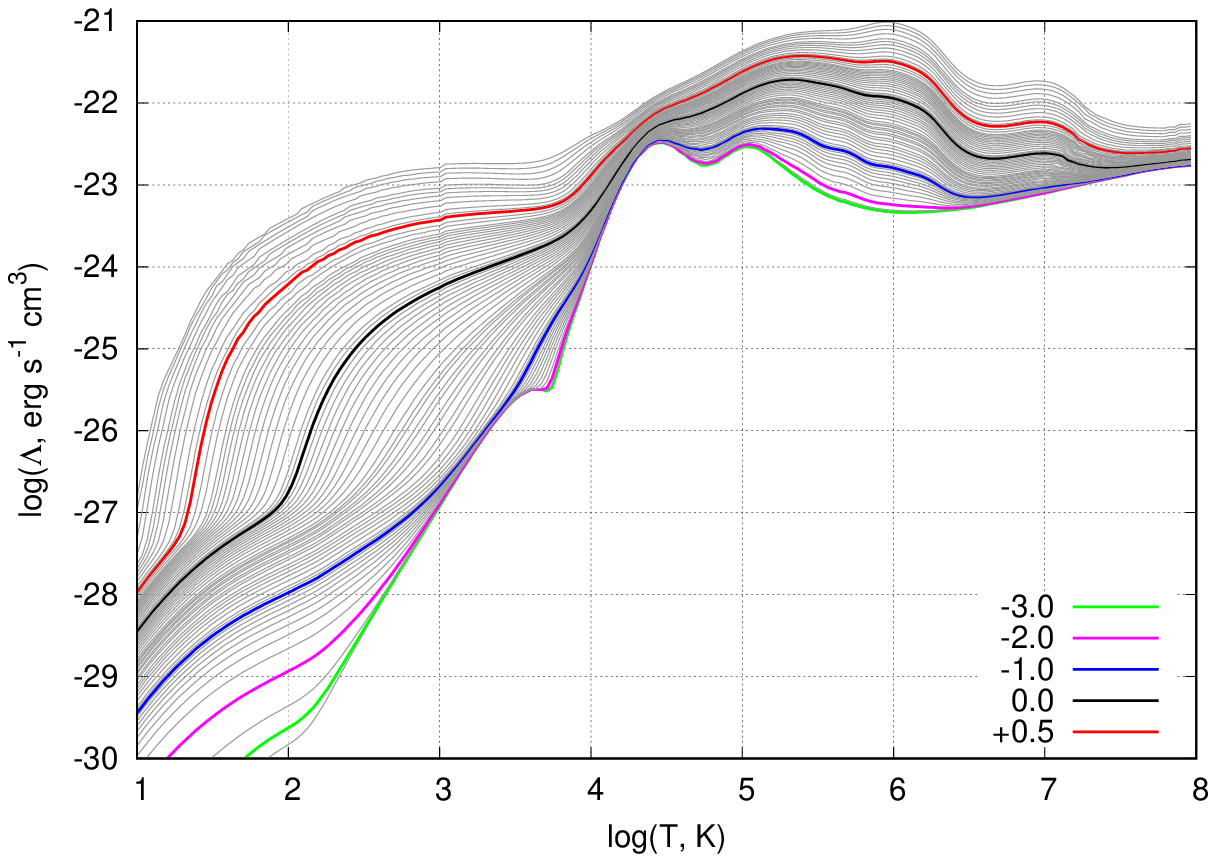}
\caption{
{
Isochoric cooling rates in the metallicity range [Z/H]$=-4..1$ with the increment $[\Delta{\rm Z/H}]$, calculated using the method described in \citet{v13}. 
}
         }
\label{fig-coolfun}
\end{figure}

In our simulations we have used a non-equilibrium cooling function $\Lambda(T,Z)$ that includes time-dependent evolution of the ionization state of the dominant coolants: He, C, N, O, Ne, Mg, Si, Fe, implemented self-consistently into cooling processes described by \citet{v13}, $Z$ is the abundance of heavy elements. In \citep{v13} calculations of the gas ionization and thermal state begin from the initial conditions at $T=10^8$ K when all dominant elements are fully ionized and then are supposed to recombine free without additional ionization and heating sources. We make use only the isochoric version of the cooling function because hydrodynamical processes relax to pressure equilibrium on relevant time scales of a Myr. In our simulations the $\Lambda(T,Z)$ function is tabulated properly, and when necessary splined between different $T$ and $Z$ corresponding to their values in a given cell and at a given time. Figure \ref{fig-coolfun} shows a graphical representation of $\Lambda(T,Z)$, the increment $[\Delta {\rm Z/H}]$ is calculated from the requierements that 

\be 
\Delta\Lambda(T,Z)={1\over\Lambda(T,z)}{\de\Lambda(T,Z)\over\de [{\rm Z/H}]} [\Delta{\rm Z/H}]\leq -0.04,
\ee
resulting in an approximate difference between subsequent values of $\Lambda(T,Z)$ by not more than 10--12\%. 

\section{Evolution of dust size distribution} \label{sputter}

At $T>10^6$ K thermal sputtering rate of a dust particle is $\dot a \simeq -3\times 10^{-18}n$ cm s$^{-1}$, $\dot a$ does not depend explicitly on dust radius $a$. Correspondingly, the continuity equation for dust size distribution is \citep[Eq. 72 in ][]{Laor1993} 
\be 
{\de n_a\over \de t}+\dot a {\de n_a\over\de a}=0,
\label{eq-sput}
\ee
with the solution $n_a(t)=f(a+a_0t/\tau_0)$, with $a_0/\tau_0\equiv 3\times 10^{-18}n$, $f(a)$ is the dust size distribution an the initial state. For the ``standard'' MRN distribution $n_a\propto a^{-3.5}$ within $a_1\leq a\leq a_2$ \citep{Mathis1977}
the distribution remains similar $n_a(t)\propto (a+a_0t/\tau_0)^{-3.5}$  with a decreasing magnitude and with the minimum size $a_a(t)=a_1+a_0t/\tau_0$.  

\end{document}